\documentclass[%
 reprint,
showpacs,preprintnumbers,
 amsmath,amssymb,
 aps,
prc,
]{revtex4-1}

\usepackage{graphicx}
\usepackage{dcolumn}
\usepackage{bm}

\begin{document}

\preprint{APS/123-QED}

\title{Hypernuclear matter in  a complete SU(3) symmetry group}

\author{Luiz L. Lopes}
\email{luiz\_kiske@yahoo.com.br}
\author{Debora P. Menezes}%

\affiliation{%
 Departamento de Fisica, CFM - Universidade Federal de Santa Catarina;  C.P. 476, CEP 88.040-900, Florian\'opolis, SC, Brasil 
}%





\begin{abstract}
 Using the well known quantum hadrodynamics (QHD), we study the effects of meson-hyperon coupling constants on the onset of
 hyperons in dense nuclear matter. We use the SU(3) symmetry group  to fix all these coupling constants, constrained to
 experimental nuclear matter results and astrophysical observations. While the discovery of  massive pulsars PSR J1614-2230 and PSR J0348+0432
 points towards a very stiff equation of state at very large densities, results from heavy ion collisions point in opposite direction for
 densities below five times the nuclear saturation density. We study some well known parametrizations and see that most of them cannot satisfy
 both types of constraints.  Indeed, although in our model we can simulate a 2.25 $M_\odot$ hyperonic neutron star, the heavy ion collision
 constraints limit the maximum mass around 2.06 $M_\odot$. 
\end{abstract}

\maketitle


\section{Introduction \label{sec1}}

The physics of nuclear matter is well understood around the nuclear saturation density, and most physical parameters are known within very little uncertainties. However, the physics of very high densities is  far from being fully understood. The discovery of  massive pulsars  PSR J1614-2230~\cite{Demo} and PSR J0348+0432~\cite{Antoniadis} indicates a very stiff equation of state (EoS) for densities above the saturation point. At such high densities, particles with some strange content can be created, since they are energetically favorable once the the Fermi energy of the nucleons becomes of the order of their rest masses. The onset of hyperons softens the EoS and reduces the possible maximum mass of the correspondent  neutron star~\cite{Glen}, which may cause a conflict between the astrophysical observations and the theoretical previsions. 

Generally models containing hyperons predict neutron stars  with masses below $1.9~M_\odot$, which would exclude the possibility of hyperons in their cores.
  Using Brueckner-Hartree-Fock (BHF) models, previous studies found maximum masses between 1.26 $M_\odot$ and 1.47 $M_\odot$~\cite{Schu,Baldo1,Baldo2,Djapo}.
 Other nonrelativistic phenomenological models increase but not enough the maximum mass to 1.8 $M_\odot$~\cite{Balberg}.
  Another possibility is to use relativistic quantum hadrodynamic (QHD) models \cite{Serot},
 where the strong interaction is simulated by the exchange of massive mesons through Yukawa potentials.
 Whilst most hypernuclear EoS predict neutron star masses  below 2 $M_\odot$ even in QHD~\cite{Glen,Haensel,Hofm,Magno,Glen3},
 a few  parametrizations succeed in describing very massive hyperonic neutron stars~\cite{Rafa,Paoli,Katayama}. 
A clever mechanism to increase the maximum mass of neutron stars with hyperons in their core (hyperonic stars) is to include an additional vector meson that mediates the hyperon-hyperon interaction~\cite{Weiss,Weiss2}. This approach has the advantage of  not affecting any of the well known nuclear properties, just pushing away the hyperon threshold and suppressing their fraction at high densities.

The major difficulty when hyperons are present is to fix the coupling constants of these particles with the mesons. It is very common in the literature the use of the hybrid SU(6) symmetry group~\cite{Pais} to fix the coupling of the hyperons with the vector mesons, while the coupling with the  scalar ones  is fixed through the hyperon potential depth. Despite the insight gained by the knowledge of the hyperon potential depth, the problem is far from solved. While the $\Lambda$ potential depth ($U_\Lambda$) is  known to be equal to -28 MeV~\cite{Glen2,Schaffner}, the $\Sigma$ and $\Xi$ potentials ($U_\Sigma$, $U_\Xi$) present a huge uncertainty and not even the  signs of these potentials are well defined~\cite{Schaffner,Dover}, although there are canonical values normally used in  modern literature ($U_\Sigma$ = +30 MeV and $U_\Xi$ = -18 MeV)~\cite{Schaffner2}. 

The use of the SU(6) parametrization for the vector mesons  along the canonical hyperon potential depths create new problems: the maximum masses usually achieved are lower than 2.0 $M_\odot$~\cite{Weiss,Rafa}, in disagreement with recent observations. An attempt to conciliate massive neutron stars with the SU(6) theory was performed in two works~\cite{Weiss,Zhao}, where the authors varied the hyperon potentials in an ad hoc way. They found that the hyperon potential has but little influence on the maximum mass. Another ad hoc approach was proposed by Glendenning et al~\cite{Glen2}, where the meson hyperon coupling was fixed so as to reproduce the $U_\Lambda$ potential
\cite{Rufa}, considering that all the hyperons couple in the same way. With this approach, a family of hyperon meson parametrization was obtained.
 Although it can describe massive neutron stars  the results are strongly model dependent, and the maximum mass can vary $ 100\%$, from 1.2 to 2.4
 solar masses~\cite{Glen}. 
Nevertheless, this approach is widely used in the literature~\cite{Glen,Magno,Paoli,Lopes2,Rafa}, and we next call it {\it Glendenning Conjecture} (GC).

Other proposals consider the inclusion of a new vector meson to mediate hyperon-hyperon interaction without affecting nuclear matter~\cite{Weiss,Weiss2}, and the break of SU(6) symmetry of the isoscalar-vector meson to a more general SU(3) flavor symmetry group~\cite{Weiss2,Myung}. The use of SU(3) instead the SU(6) symmetry produces very massive hyperonic neutron stars, with maximum masses as high as 2.3 $M_\odot$~\cite{Weiss2}.
The main problem  related to the previous SU(3) symmetry based models is that while the isoscalar-vector mesons are allowed to break the SU(6) symmetry, the isovector-vector meson is forced to  obey it, what sounds rather artificial. The  scalar meson coupling is even poorly obtained, since it is fixed from a pure phenomenological basis. Moreover, results of the last decade in heavy ions collision (HIC) have shown that although the EoS has to be very stiff at large densities, it should be soft for densities below five times the nuclear saturation point~\cite{JSB,Daniel}, what is in disagreement with EoS that produce  high mass neutron stars.
	
In this work, we study a QHD based model with three  parametrization: GM1, GM3~\cite{Glen2} and NL3~\cite{Lala} in the traditional $\sigma\omega\rho$ and in the more exotic $\sigma\omega\rho\phi$ models~\cite{Weiss,Weiss2}. We  propose new parametrization families in such a way that all  hyperon meson couplings are in accordance with the SU(3) symmetry group instead of only the isoscalar-vector ones. We do that tying the parameters of the scalar meson group with the vector ones, in such a way to describe the well known $U_\Lambda$ = -28 MeV, and let  the uncertain $U_\Sigma$ and $U_\Xi$ be determined only by the symmetry group properties, without any other phenomenological input. We also use one of  the GC parametrization in the standard $\sigma\omega\rho$ model to compare the differences arising from a full symmetric model with respect to a full phenomenological one. Then, we calculate some physical quantities as particle fractions, strangeness fraction and speed of sound in the hypernuclear medium, since  they may be important to investigate  quark-hadron phase transitions~\cite{Prakash}.
In order to validate our proposal we compare our results with those obtained from HIC and astrophysical observations. 

This paper is organized as follows: in section \ref{sec2} we discuss the QHD formalism, calculate the EoS and the meson field equations.
 In section \ref{sec3} we display our choice of parametrizations and some of the physical quantities they foresee for  nuclear matter.
 We also display all the hyperon meson coupling constants  obtained from the SU(3)
 symmetry and the GC approach and the related previsions for the hyperon potential depths. 
In section \ref{sec4}  we show the results of several parametrization sets and discuss some observational, experimental and theoretical constraints.
 The conclusions are presented in section \ref{sec5} and the calculations of the hyperon-meson coupling constants within a complete SU(3)
 symmetry are developed in the Appendix.

\section{Formalism \label{sec2}}

We use an extended version of the relativistic QHD~\cite{Serot}, whose Lagrangian density reads:
\begin{widetext}
\begin{eqnarray}
\mathcal{L}_{QHD} = \sum_B \bar{\psi}_B[\gamma^\mu(i\partial_\mu  - g_{BB\omega}\omega_\mu  - g_{BB\phi}\phi_\mu - g_{BB\rho} \frac{1}{2}\vec{\tau} \cdot \vec{\rho}_\mu)
- (m_B - g_{BB\sigma}\sigma)]\psi_B  -U(\sigma) +   \nonumber   \\
  + \frac{1}{2}(\partial_\mu \sigma \partial^\mu \sigma - m_s^2\sigma^2) + \frac{1}{4}\Omega^{\mu \nu}\Omega_{\mu \nu} + \frac{1}{2} m_v^2 \omega_\mu \omega^\mu - \frac{1}{4}\Phi^{\mu \nu}\Phi_{\mu \nu} 
+ \frac{1}{2}m_\phi^2 \phi_\mu \phi^\mu + \frac{1}{2} m_\rho^2 \vec{\rho}_\mu \cdot \vec{\rho}^{ \; \mu} - \frac{1}{4}\bf{P}^{\mu \nu} \cdot \bf{P}_{\mu \nu}  , \label{s1} 
\end{eqnarray}
\end{widetext}
where the sum in $B$ stands just for the nucleons or for all the baryon octet,
{depending on our choice for the star consitituents,   
 $\psi_B$  are the baryonic  Dirac fields, and $\sigma$, $\omega_\mu$, $\phi_\mu$ and $\vec{\rho}_\mu$ are the mesonic fields. The $g's$ are the Yukawa coupling constants that simulate the strong interaction, $m_B$ is the mass of the baryon $B$, $m_s$, $m_v$, $m_\phi$ and $m_\rho$ are the masses of the $\sigma$, $\omega$, $\phi$ and $\rho$ mesons respectively. The antisymmetric mesonic field strength tensors are given by their usual expressions as presented in~\cite{Glen}.  The $U(\sigma)$ is the self-interaction term introduced in ref.~\cite{Boguta} to reproduce some of the saturation properties of the nuclear matter and is given by:
 
 \begin{equation}
U(\sigma) =  \frac{1}{3!}\kappa \sigma^3 + \frac{1}{4!}\lambda \sigma^{4} \label{s2} .
\end{equation}
 
 Finally, $\vec{\tau}$ are the Pauli matrices. In order to describe a neutral, chemically stable hypernuclear matter, we add leptons as free Fermi gases:
 
 \begin{equation}
 \mathcal{L}_{lep} = \sum_l \bar{\psi}_l [i\gamma^\mu\partial_\mu -m_l]\psi_l , \label{s3}
 \end{equation}
 where the sum runs over the two lightest leptons ($e$ and $\mu$).
 
The nucleon masses are assumed to be  $N=939$ MeV, the $\Sigma$ triplet masses are 1193 MeV, the $\Lambda^0$ mass is 1116 MeV, and the $\Xi$ doublet masses are 1318 MeV.
 The electron and muon masses are 0.511 MeV and 105.6 MeV respectively. The vector meson masses are 783 MeV for the $\omega$ and 1020 MeV for the $\phi$. With  GM1 and GM3 parametrizations, we have $770$ MeV for the $\rho$, while with the NL3, the $\rho$ mass is $763$ MeV.
 All these masses are the physical ones, corresponding to values close to those found 
experimentally. The scalar meson $\sigma$ may be regarded as the $\epsilon(760)$~\cite{Swart5,Greiner,Rijken} with a fictitious mass
 of 512 MeV for GM1 and GM3, and 508 MeV for NL3, to agree with the parametrizations proposed in~\cite{Glen2,Lala}.

To solve the equations of motion, we use the mean field approximation (MFA), where the meson fields are replaced by their expectation values, i.e:  $\sigma$ $\to$ $\left < \sigma \right >$ = $\sigma_0$,   $\omega^\mu$ $\to$ $\delta_{0 \mu}\left <\omega^\mu  \right >$ = $\omega_{0}$  and   $\rho^\mu$ $\to$ $\delta_{0 \mu}\left <\rho^\mu  \right >$ = $\rho_{0}$.
 The MFA gives us the following eigenvalue for the baryon energy~\cite{Glen}:

\begin{equation}
E_B = \sqrt{k^2 + M^{*2}_B} + g_{BB\omega}\omega_0 + g_{BB\rho} \frac{\tau_3}{2}  \rho_0 ,  \label{s4}
\end{equation}
where $M_B^*$ is the baryon effective mass: $M^{*}_B$ $ \dot{=} $ $  m_B - g_{BB\sigma}\sigma_0$. 

For the leptons, the energy eigenvalues are those of the free Fermi gas:
\begin{equation}
\quad E_l = \sqrt{k^2 + m^{2}_l} , \label{s5}
\end{equation}
and the meson fields become:

 \begin{equation}
\omega_0  =\sum_B \frac{g_{BB\omega}}{m_v^2} n_B , \label{s6}
\end{equation}

\begin{equation}
\phi_0  = \sum_B \frac{g_{BB\phi}}{m_\phi^2} n_B,  \label{s7}
\end{equation}

\begin{equation}
\rho_0  = \sum_B \frac{g_{BB\rho}}{m_\rho^2} \frac{\tau_{3}}{2} n_B,  \label{s8}
\end{equation}

\begin{equation}
\sigma_0 =  \sum_B \frac{g_{BB\sigma}}{m_s^2}  n_{SB} - \frac{1}{2}\frac{\kappa}{m_s^2}\sigma_0^2 -\frac{1}{6}\frac{\lambda}{m_s^2} \sigma_0^3 , \label{s9}
\end{equation}
where $n_{SB}$ is the scalar density  and $n_{B}$ is the number  density of the baryon $B$:

\begin{eqnarray}
\quad n_{SB} =  \int_0^{k_{fB}} \frac{M^*}{\sqrt{k^2 + M^{*2}}} \frac{k^2}{\pi^2} dk  \quad  , \nonumber
 \\ n_B = \frac{k_{fB}^3}{3\pi^2}, \quad \mbox{and}  \quad  n= \sum_B n_B 
\label{s10}
\end{eqnarray}

To describe the properties of the hypernuclear matter, we calculate the EoS from statistical mechanics~\cite{Greiner2}. The baryons and leptons, being fermions, obey the Fermi-Dirac distribution. In order to compare our results with experimental and 
observational constraints, we next study nuclear and stellar systems at zero temperature.
In this case the Fermi-Dirac distribution becomes the Heaviside step function. The  energy densities of  baryons, leptons and  mesons (which are bosons) read:

\begin{equation}
\epsilon_B =  \frac{1}{\pi^2} \sum_B \int_0^{k_f} \sqrt{k^2 + M^{*2}_B} k^2 dk , \label{s11}
\end{equation}
\begin{equation}
\epsilon_l = \frac{1}{\pi^2} \sum_l \int_0^{k_f} \sqrt{k^2 + m^{2}_l} k^2 dk , \label{s12}
\end{equation}
\begin{equation}
\epsilon_m =  \frac{1}{2}\bigg ( m_s^2\sigma_0^2 + m_v^2\omega_0^2 +m_\phi^2\phi_0^2 + m_\rho^2\rho_0^2 \bigg ) + U(\sigma) , \label{s13}
\end{equation}
where $k_f$ is the Fermi momentum, and we have already used the fact that the fermions have degeneracy  equal to 2. The total energy density is the sum of the partial ones:

\begin{equation}
\epsilon =   \epsilon_B +   \epsilon_l +  \epsilon_m ,\label{s14}
\end{equation}
and the pressure is calculated via thermodynamic relations:

\begin{equation}
P = \sum_f \mu_f n_f - \epsilon , \label{s15}
\end{equation}
where the sum runs over all the fermions ($f = B,l$) and $\mu$ is the chemical potential, which
corresponds exactly to the energy eigenvalue at $T=0$.

\section{Model parameters \label{sec3}}

We use three standard QHD parametrization: GM1, GM3~\cite{Glen2,Paoli} and NL3~\cite{Lala} to describe 
 five input parameters: nuclear saturation density, $n_0$,  binding energy per baryon $B/A$, effective nucleon mass $M^*$, nuclear compression modulus $K$
 and the symmetry energy coefficient $S_0$. Table \ref{T1} resumes the parametrizations and the bulk nuclear matter values they generate.

\begin{table}[ht]
\begin{center}
\begin{tabular}{|c||c|c|c|}
\hline 
  & GM1 &  GM3 & NL3  \\
 \hline
 $g_{NN\sigma}$ & 8.910  & 8.175 & 10.217  \\
 \hline
  $g_{NN\omega}$ & 10.610 & 8.712 & 12.868 \\
 \hline
  $g_{NN\phi}$ & 0.0 & 0.0 & 0.0 \\
 \hline
  $g_{NN\rho}$ & 8.196 & 8.259 & 8.948 \\
 \hline
$\kappa/M_N$ & 0.005894 & 0.017318 & 0.0041014  \\
\hline
$\lambda$ &  -0.006426 & -0.014526 & -0.015921 \\
\hline  
\hline
$n_0$ $(fm^{-3})$ & 0.153 & 0.153 & 0.148 \\
\hline
$M^{*}/M$ & 0.70 & 0.78 & 0.60 \\
\hline
K $(MeV)$ & 300 & 240 & 272 \\
\hline
$S_0$ $(MeV)$ & 32.5 & 32.5 & 37.4 \\
\hline
$B/A$  $(MeV)$ & -16.3 & -16.3 & -16.3 \\ 
\hline
\end{tabular}
 
\caption{Parameters and physical quantities for GM1, GM3 and NL3 models.} \label{T1}
\end{center}
\end{table}

The $\phi$ meson, which carries strangeness, is generally disregarded.  The models where it is not present are the traditional $\sigma\omega\rho$ models. When the $\phi$ is present, we have chosen to call the models as $\sigma\omega\rho\phi$ models
~\cite{Weiss,Weiss2,Mishustin1996,Greiner2002,Rafael2008}.

In order to fix the hyperon meson couplings, we use two different approaches. 
The first one relies on a pure phenomenological basis and we call it GC.
Within the GC parametrization, the $\phi$ meson is never present. 
It assumes that ~\cite{Glen2}:
\begin{equation}
\frac{g_{YY\sigma}}{g_{NN\sigma}} = 0.7, \quad \frac{g_{YY\omega}}{g_{NN\omega}} = \chi_\omega  \quad \frac{g_{YY\rho}}{g_{NN\rho}} = \frac{I_{3B}}{I_{3N}} \chi_\rho , \label{s16}
\end{equation}
where $\chi_\omega = \chi_\rho$ has the values  0.783 in GM1, 0.8 in GM3 and 0.772 in NL3 in order to describe the well known $U_\Lambda = -28$ MeV. The hyperon potential depth is defined as~\cite{Glen,Weiss}:
\begin{equation}
U_Y =  g_{YY\omega}\omega_0 + g_{YY\phi}\phi_0 - g_{YY\sigma}\sigma_0 . \label{s17}
\end{equation}

 Note that within the GC parametrization the $\rho$ meson always couples 
to the isospin projection $I_3$. Nevertheless, the value of $\chi_\rho$ is 
completely arbitrary~\cite{Glen}.

 Another possible approach is the assumption of a complete SU(3) symmetry group theory to determine the coupling constants of baryons with all
 mesons~\footnotetext{1}[A detailed calculation is present in the Appendix.]. We also assume $z_v =\sqrt{6}$ and $\theta_v = 35.264$ in agreement with the SU(6)
 parametrization of the vector mesons, and a {\it near SU(6)}, with $z_s = \frac{8}{9}\sqrt{6}$ and $\theta_s = 35.264$ for the scalar mesons.
 Within this complete SU(3) symmetry group model, the  hyperon-meson couplings now become:
 \begin{eqnarray}
\frac{g_{\Lambda\Lambda\omega}}{g_{NN\omega}} = \frac{4 + 2\alpha_v}{5 + 4\alpha_v}, \quad \frac{g_{\Sigma\Sigma\omega}}{g_{NN\omega}}  =
 \frac{8 - 2\alpha_v}{5 + 4\alpha_v}, \nonumber \\ \quad \frac{g_{\Xi\Xi\omega}}{g_{NN\omega}} = \frac{5 - 2\alpha_v}{5 + 4\alpha_v}, \label{s18}
\end{eqnarray}
for the meson $\omega$, and for the meson $\phi$, when it is present:
\begin{eqnarray}
\frac{g_{NN\phi}}{g_{NN\omega}} = \sqrt{2} \cdot \bigg (\frac{4\alpha_v - 4}{5 + 4\alpha_v} \bigg ), \quad
 \frac{g_{\Lambda\Lambda\phi}}{g_{NN\omega}} = \sqrt{2} \cdot \bigg (\frac{2\alpha_v - 5}{5 + 4\alpha_v} \bigg ),  \nonumber \\
\frac{g_{\Sigma\Sigma\phi}}{g_{\Lambda\Lambda\omega}} = \sqrt{2} \cdot \bigg (\frac{-2\alpha_v - 1}{5 + 4\alpha_v} \bigg ),
 \quad \frac{g_{\Xi\Xi\phi}}{g_{NN\omega}} = \sqrt{2} \cdot \bigg (\frac{-2\alpha_v - 4}{5 + 4\alpha_v} \bigg ), \nonumber \\ \label{s19}
\end{eqnarray}
In the traditional $\sigma\omega\rho$ model we consider that all $g_{BB\phi}$ are equal to zero.

For the meson $\rho$:
\begin{equation}
\frac{g_{\Sigma\Sigma\rho}}{g_{NN\rho}} = 2\alpha_v, \quad \frac{g_{\Xi\Xi\rho}}{g_{NN\rho}} = -(1 -2\alpha_v), \quad \frac{g_{\Lambda\Lambda\rho}}{g_{NN\rho}} = 0, \label{s20}
\end{equation}
and for the $\sigma$:

\begin{eqnarray}
\frac{g_{\Lambda\Lambda\sigma}}{g_{NN\sigma}} = \frac{10+ 6\alpha_s}{13 + 12\alpha_s}, \quad \frac{g_{\Sigma\Sigma\sigma}}{g_{NN\sigma}}
  = \frac{22 - 6\alpha_s}{13 + 12\alpha_s}, \nonumber \\ \frac{g_{\Xi\Xi\sigma}}{g_{NN\sigma}} = \frac{13 - 6\alpha_s}{13 + 12\alpha_s}. \label{s21}
\end{eqnarray}

When we set $\alpha_v =1$, we recover the SU(6) parametrization for the vector mesons. In this case, the $\omega$ meson couples to hypercharge, while the $\rho$ meson couples to isospin, as proposed by Sakurai~\cite{Sakurai}. When $\alpha \neq$ 1, the $\phi$ meson couples to the nucleon in the $\sigma\omega\rho\phi$ model. To make sure that the nuclear matter properties are not affected, we reparametrize the $g_{NN\omega}$  as follows:

\begin{equation}
g_{NN\omega}\omega_0 \to \tilde{g}_{NN\omega}\omega_0 + g_{NN\phi}\phi_\phi,  
\label{s22}
\end{equation}
where the left side of Eq. (\ref{s22}) is related to the $\sigma\omega\phi$ model,
 while the right side to the $\sigma\omega\rho\phi$ model. Now, from Eqs.(\ref{s6}) and (\ref{s7}):

\begin{eqnarray}
 \quad g_{NN\omega}\sum_B  \frac{g_{BB\omega}}{m_v^2} n_B \; \equiv \; \nonumber \\ \tilde{g}_{NN\omega}\sum_B\frac{\tilde{g}_{BB\omega}}{m_v^2} n_B + g_{NN\phi}\sum_B \frac{g_{BB\phi}}{m_\phi^2} n_B . \label{s23}
\end{eqnarray}

However, since hyperons are not present at the nuclear saturation density, the sum runs only over the nucleons. Also, $g_{NN\phi}$ depends on the independent $g_{NN\omega}$ from Eq. (\ref{s19}). Therefore we rewrite  Eq. (\ref{s23}) as:
\begin{equation}
\frac{g_{NN\omega}^2}{m_v^2}  \; \equiv \;  \frac{\tilde{g}_{NN\omega}^2}{m_v^2}  + 2  \bigg (\frac{4\alpha_v - 4}{5 + 4\alpha_v} \bigg )^2 \frac{\tilde{g}_{NN\omega}^2}{m_\phi^2} .  \label{s24}
\end{equation}
 
 Notice that for each $\alpha_v$, $\tilde{g}_{NN\omega}$ in $\sigma\omega\rho\phi$ model on the right side of Eq. (\ref{s24}) is
  determined in such a way that the right side of that equation  reproduces the same value of the left side in the traditional $\sigma\omega\rho$ model.
 We call the parametrizations of the $\sigma\omega\rho\phi$ model as ``Like-Model, LM" (i.e, GM1LM, GM3LM, etc) since they predict
 the same quantities for the nuclear saturation as the original models, although with different parameters. As already pointed out in \cite{Weiss2},
 the unusual $N-\phi$  coupling that arises when we break the SU(6) symmetry in the LM  are in accordance with the huge strange quark condensate in the nucleon found in lattice gauge
 simulations~\cite{Dong,Ellis,Eng}.

 \subsection{Hyperon-meson couplings and potential depths}
 
 According to the calculations developed in the Appendix, we have a priori just two free parameters, $\alpha_v$ and $\alpha_s$. Now we proceed as follow: we give arbitrary values to  $\alpha_v$ varying from 1 to 0, and impose that $\alpha_s$ assumes the value that keeps $U_\Lambda$ = -28 MeV. The uncertain $U_\Sigma$ and $U_\Xi$  are then determined only by symmetry properties, without any other phenomenological input. 
We also display the values for $g_{NN\omega}$, that change within the LM. The results are presented from Table \ref{T2}  to Table \ref{T5}.

\begin{table}[ht]
\begin{tabular}{|c||c|c|c|c|}
\hline
 $\alpha_v = 1$ & $\alpha_s = 1.568$ & $g_{NN\omega} = 10.610$  & $U_\Sigma = +32$ & $U_\Xi = +40$   \\
 \hline
\hline
 $\alpha_v = 0.75$ & $\alpha_s = 1.251$ & $g_{NN\omega} = 10.610$  & $U_\Sigma = +29$ & $U_\Xi = +39$   \\
 \hline
\hline
 $\alpha_v = 0.50$ & $\alpha_s = 0.9007$ & $g_{NN\omega} = 10.610$ & $U_\Sigma = +19$ & $U_\Xi = +33$   \\
 \hline
\hline
 $\alpha_v = 0.25$ & $\alpha_s = 0.5230$ & $g_{NN\omega} = 10.610$  & $U_\Sigma = +11$ & $U_\Xi = +29$   \\
 \hline
\hline
\hline
 $\alpha_v = 0.0$ & $\alpha_s = 0.2859$ & $g_{NN\omega} = 10.610$  & $U_\Sigma = -2$ & $U_\Xi = +22$   \\
 \hline
\hline
\hline
GC & - & $g_{NN\omega} = 10.610$ & $U_\Lambda = -28$  & $U_\Xi = -28$   \\
 \hline
\end{tabular}
 \caption{Family of parametrizations and hyperon potential depths for GM1.}\label{T2}
 \end{table}

 We see that for the SU(6) symmetry ($\alpha_v = 1)$, the $\Xi$ potential arises naturally as strongly repulsive, in accordance with
 what was suggest in  ref.~\cite{Weiss}, but  from  an ad hoc way. Also,  for the  SU(6) parametrization, we obtain $\alpha_s = 1.568$ with GM1
 and GM1LM. This value is very close to $1.496$ found in ref.~\cite{Swart5}.
 
\begin{table}[ht]
\begin{tabular}{|c||c|c|c|c|}
\hline
 $\alpha_v = 1$ & $\alpha_s = 1.678$ & $g_{NN\omega} = 8.712$  & $U_\Sigma = +22$ & $U_\Xi = +29$   \\
 \hline
\hline
 $\alpha_v = 0.75$ & $\alpha_s = 1.345$ & $g_{NN\omega} = 8.712$  & $U_\Sigma = +19$ & $U_\Xi = +28$   \\
 \hline
\hline
 $\alpha_v = 0.50$ & $\alpha_s = 1.012$ & $g_{NN\omega} = 8.712$  & $U_\Sigma = +19$ & $U_\Xi = +33$   \\
 \hline
\hline
 $\alpha_v = 0.25$ & $\alpha_s = 0.6889$ & $g_{NN\omega} = 8.712$  & $U_\Sigma = +8$ & $U_\Xi = +23$   \\
 \hline
\hline
\hline
 $\alpha_v = 0.0$ & $\alpha_s = 0.3763$ & $g_{NN\omega} = 8.712$  & $U_\Sigma = +0.2$ & $U_\Xi = +18$   \\
 \hline
\hline
\hline
GC & - & $g_{NN\omega} = 8.712$ & $U_\Lambda = -28$  & $U_\Xi = -28$   \\
 \hline
\end{tabular}
 \caption{Family of parametrizations and hyperon potential depths for GM3.}\label{T3}
 \end{table}

 \begin{table}[ht]
\begin{tabular}{|c||c|c|c|c|}
\hline
 $\alpha_v = 1$ & $\alpha_s = 1.568$ & $g_{NN\omega} = 10.610$  & $U_\Sigma = +32$ & $U_\Xi = +40$   \\
 \hline
\hline
 $\alpha_v = 0.75$ & $\alpha_s = 1.231$ & $g_{NN\omega} = 10.514$ & $U_\Sigma = +25$ & $U_\Xi = +39$   \\
 \hline
\hline
 $\alpha_v = 0.50$ & $\alpha_s = 0.8229$ & $g_{NN\omega} = 10.133$  & $U_\Sigma = -5$ & $U_\Xi = +32$   \\
 \hline
\hline
 $\alpha_v = 0.25$ & $\alpha_s = 0.6889$ & $g_{NN\omega} = 9.324$  & $U_\Sigma = -157$ & $U_\Xi = -40$   \\
 \hline
\hline
\end{tabular}
 \caption{Family of parametrizations and hyperon potential depths for GM1LM.}\label{T4}
 \end{table} 
  \begin{table}[ht]
\begin{tabular}{|c||c|c|c|c|}
\hline
 $\alpha_v = 1$ & $\alpha_s = 1.678$ & $g_{NN\omega} = 8.712$  & $U_\Sigma = +22$ & $U_\Xi = +29$   \\
 \hline
\hline
 $\alpha_v = 0.75$ & $\alpha_s = 1.367$ & $g_{NN\omega} = 8.633$  & $U_\Sigma = +20$ & $U_\Xi = +31$   \\
 \hline
\hline
 $\alpha_v = 0.50$ & $\alpha_s = 0.9281$ & $g_{NN\omega} = 8.320$  & $U_\Sigma = -2$ & $U_\Xi = +25$   \\
 \hline
\hline
 $\alpha_v = 0.25$ & $\alpha_s = 0.2775$ & $g_{NN\omega} = 7.182$  & $U_\Sigma = -104$ & $U_\Xi = -23$   \\
 \hline
\hline
\end{tabular}
 \caption{Family of parametrizations and hyperon potential depths for GM3LM.}\label{T5}
 \end{table}

 When $\alpha_v = 0.25$ with GM1LM and GM3LM, the $U_\Sigma$ becomes too attractive,
 (-157 MeV with GM1LM and -104 MeV with GM3LM). That  happens because when  

 \noindent we reduce the $\alpha_v$, the $\Sigma-\sigma$
 interaction becomes more and more important, with the increase of $g_{\Sigma\Sigma\sigma}$.
 Moreover the repulsive channel, which is the combination of  $\Sigma-\omega$ and  $\Sigma-\phi$ interactions, becomes smaller for smaller $\alpha_v$.
 These two combined factors made a stunning attractive $\Sigma$ potential.
 We show the effect of such attractive potential in Fig.~\ref{F2} and Fig.~\ref{F3} and, since this fact is in disagreement with the experience,
  these parametrizations are no longer used.
  
 We do  not study the effects of hyperons with the  NL3 parametrization, because as we see next ,
 this parametrization is in disagreement with all experimental constraints used in this work.

\section{Results and constraints \label{sec4}}

We analyze the hypernuclear matter subject to 
generalized beta equilibrium and electrically neutral conditions. 
These conditions imply that:

\begin{equation}
\mu_i = \mu_n -e_i\mu_i, \quad \mu_e =\mu_\mu, \quad \sum_B n_Be_B + \sum_l e_l+\mu_l = 0, \label{s25}
\end{equation}
where $\mu_i$ and $e_i$ are the chemical potential and the electrical charge of the $ith$ baryon respectively while $n_B$ and $n_l$ are the number densities of the baryons and leptons. Notice that in all figures and tables, the SU(6) choice of parameters refers to the case when $\alpha_v=1$.
The fraction of particles is defined as $Y_i = n_i/n$ and plot them  for the GC  and several values of $\alpha_v$ with the GM1 parametrization in Fig.~\ref{F1} 
\begin{figure*}[ht]
\begin{tabular}{cc}
\includegraphics[width=5.6cm,height=6.2cm,angle=270]{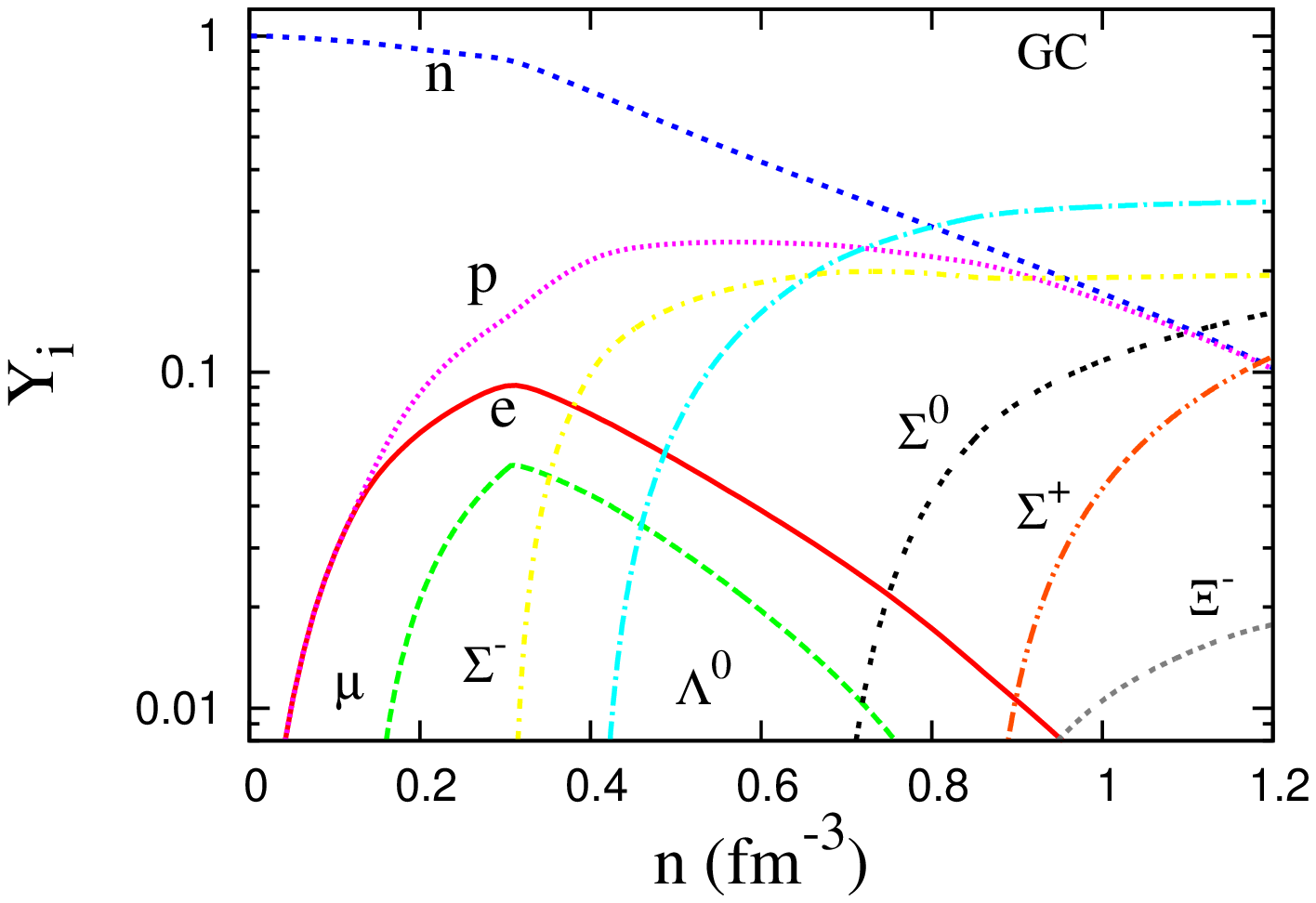} &
\includegraphics[width=5.6cm,height=6.2cm,angle=270]{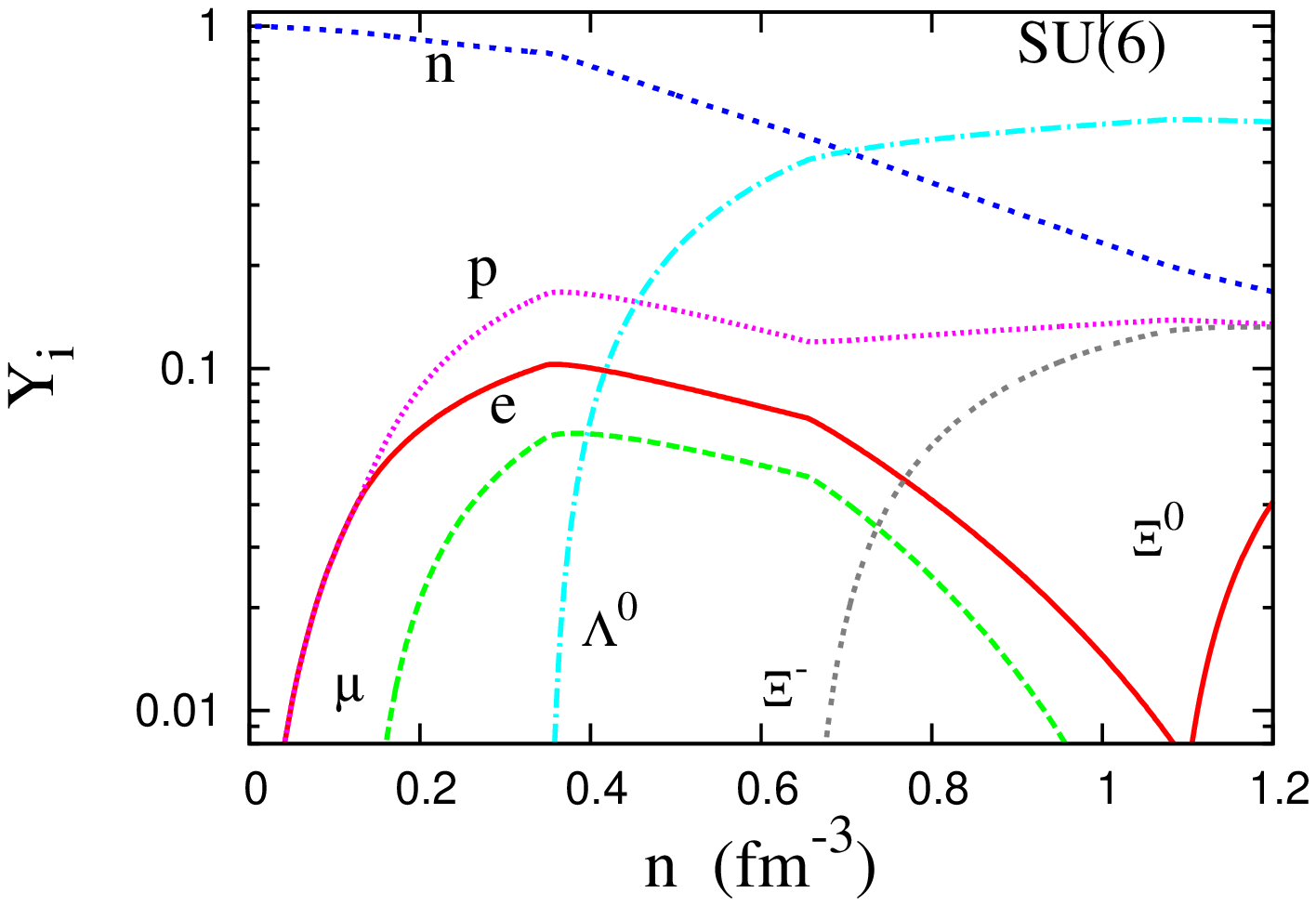} \\
\includegraphics[width=5.6cm,height=6.2cm,angle=270]{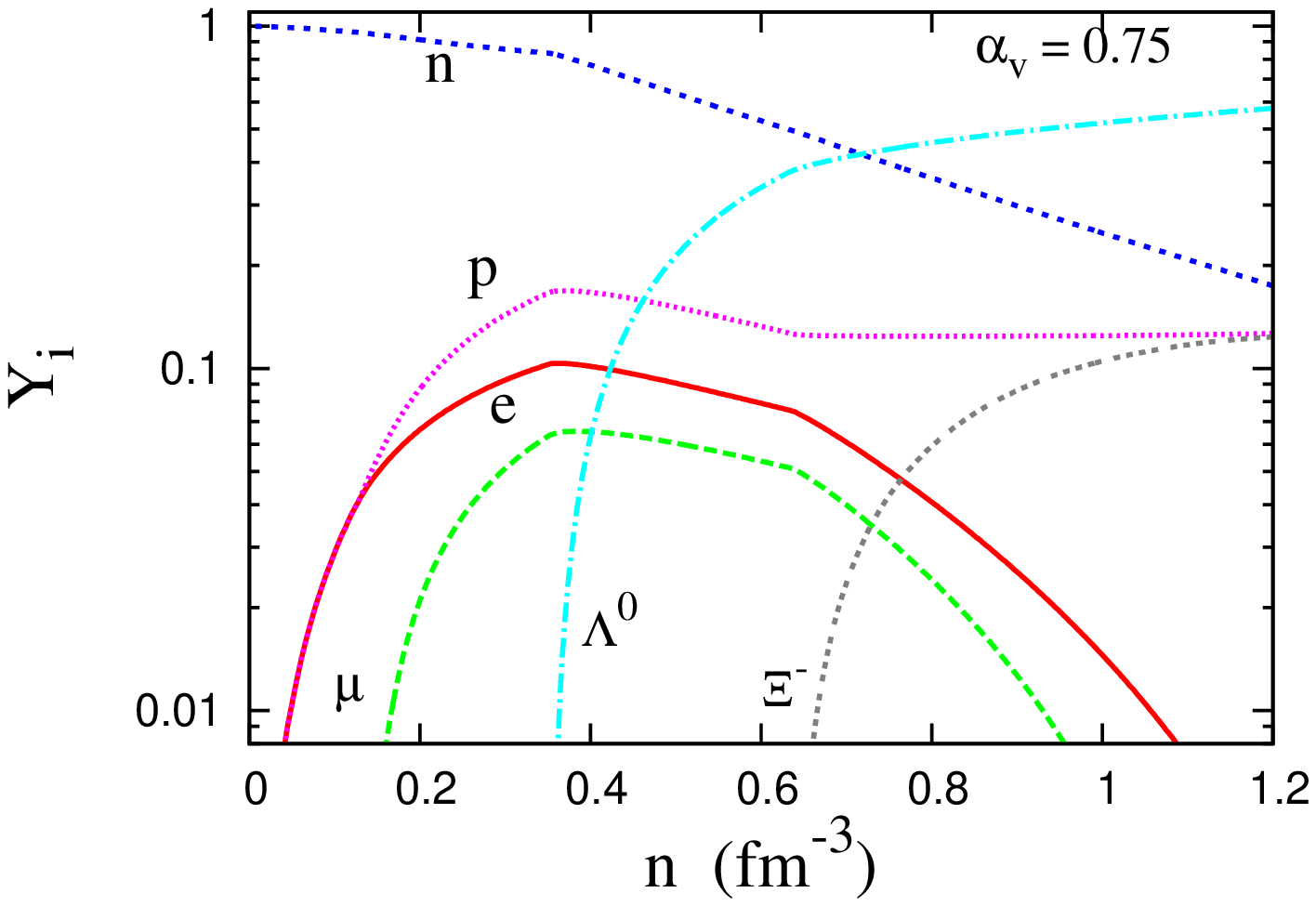} &
\includegraphics[width=5.6cm,height=6.2cm,angle=270]{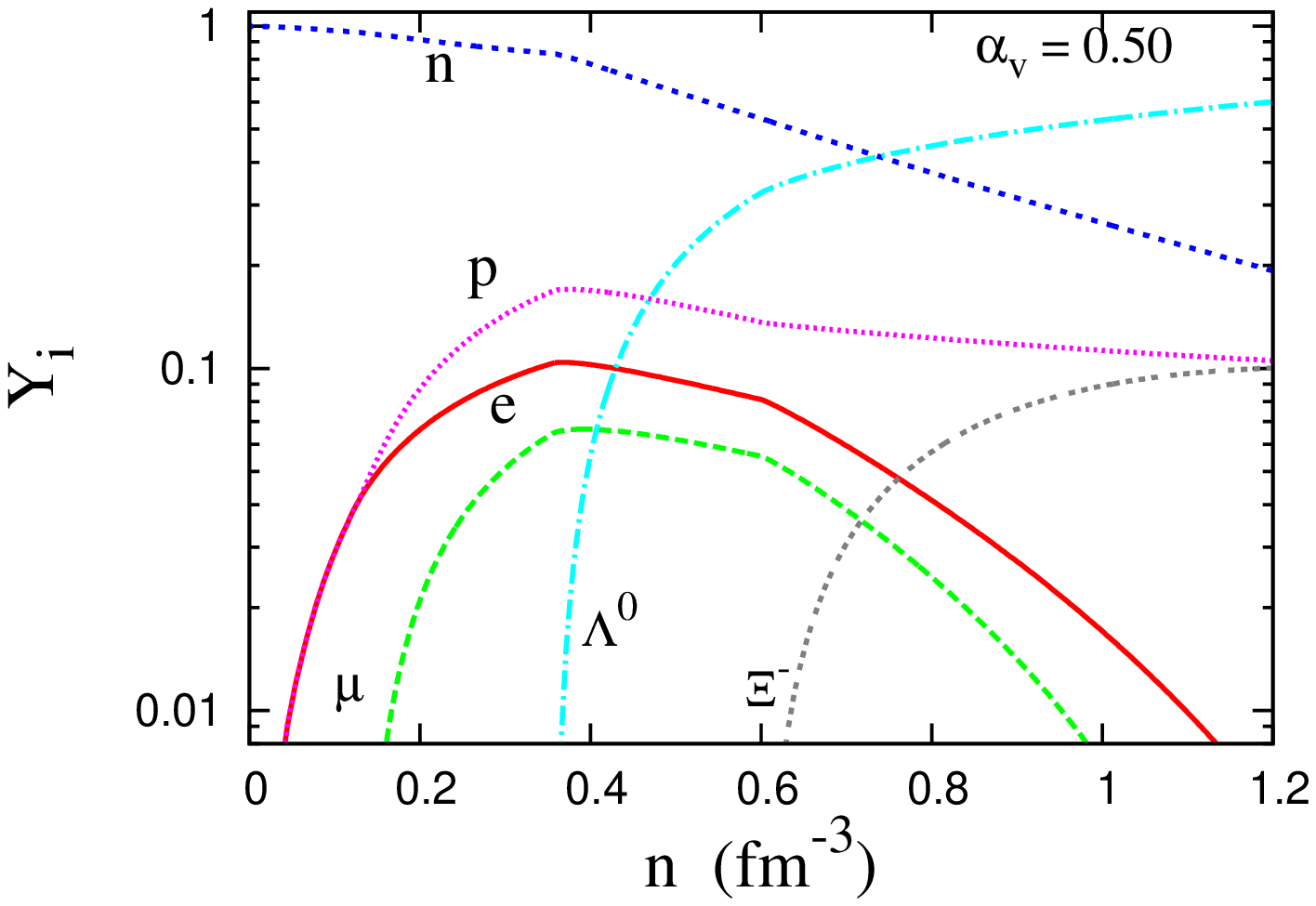} \\
\includegraphics[width=5.6cm,height=6.2cm,angle=270]{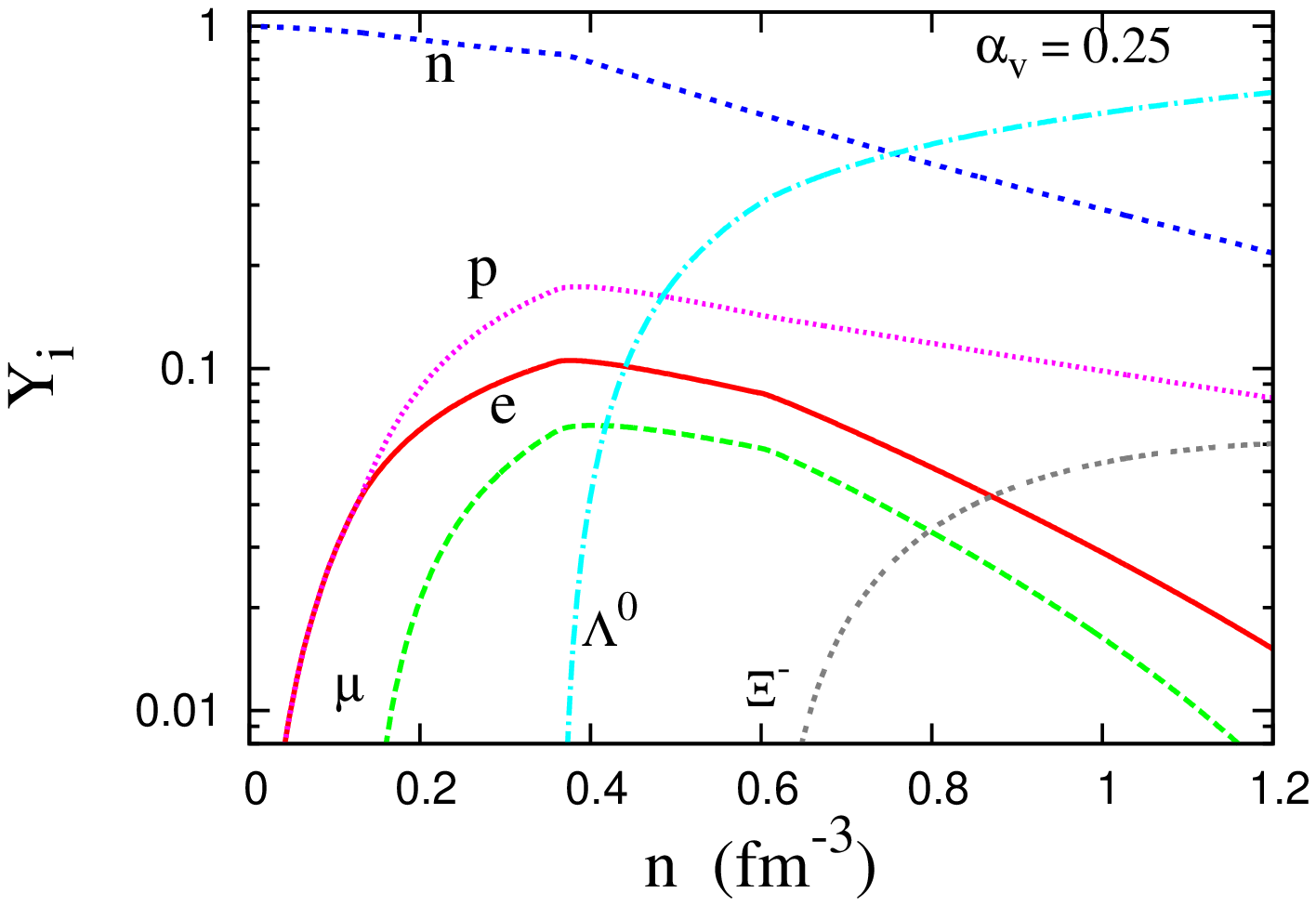} &
\includegraphics[width=5.6cm,height=6.2cm,angle=270]{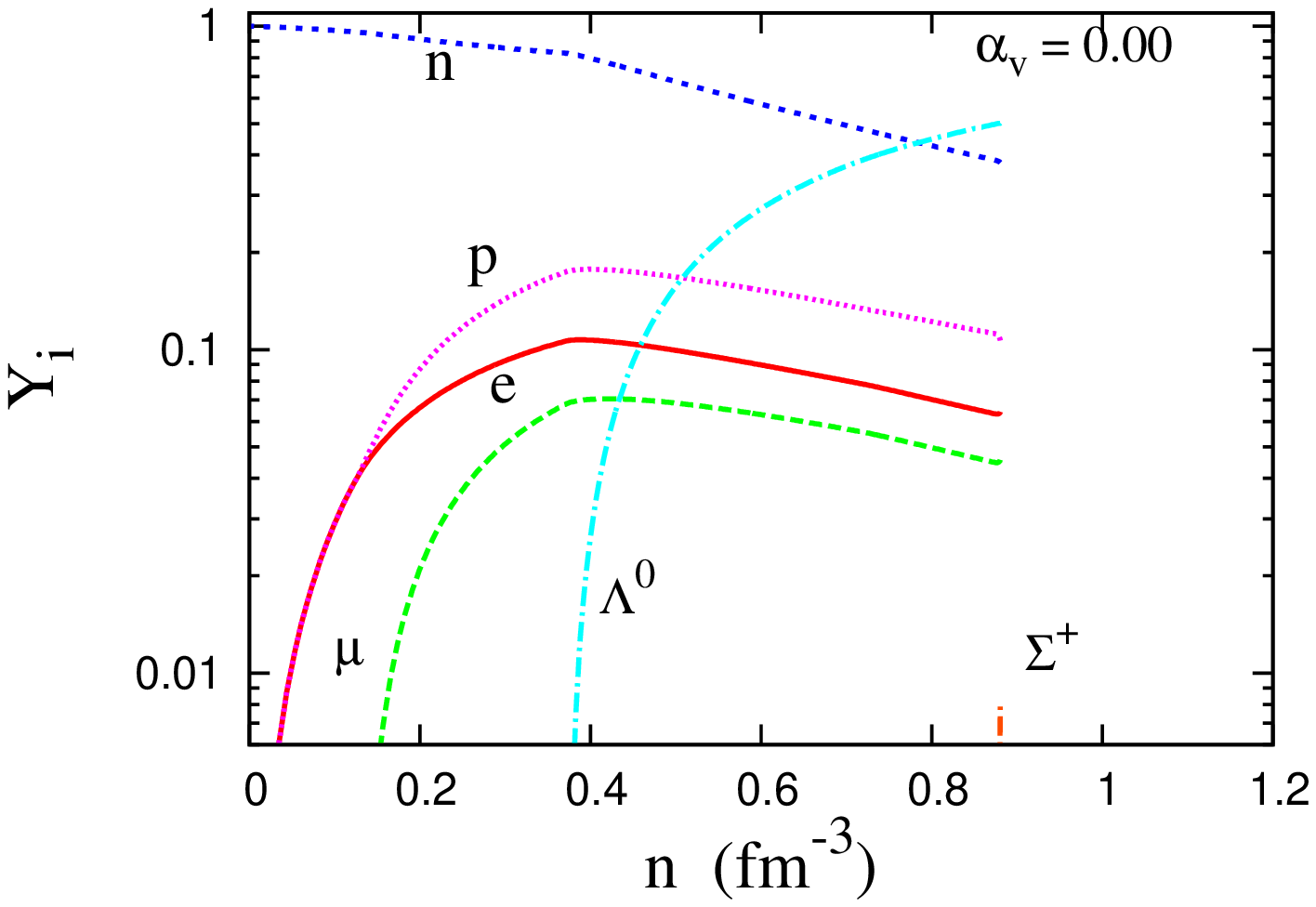} \\
\end{tabular}
\caption{Fraction of particles $Y_i$, for GC and different values of $\alpha_v$ with the  GM1 parametrization.
$\alpha_v=1$ refers to the SU(6) usual choice of couplings.} \label{F1}
\end{figure*}

 We can see that the GC parametrization differs from all those consistent with the SU(3) symmetry group, since the first hyperon that appears  with the GC is
\noindent the $\Sigma^-$ which  is absent for all SU(3) parametrizations. Indeed, while all the  $\Sigma$ triplet is present with the GC, this is not the case with the SU(3) choice of couplings, except for the subtle onset of the $\Sigma^+$ in $\alpha_v = 0.0$ at large densities. With the  SU(3) symmetry, from  $\alpha_v = 1.0$, which reproduces the usual SU(6) symmetry choice of parameters, to $\alpha_v = 0.0$, we see that the hyperons are more and more suppressed at high densities due to the increase of the repulsive vector channel. Within SU(6) there are three different hyperons in the composition of the hypernuclear matter: $\Lambda^0$, $\Xi^-$, and $\Xi^0$. When the value of $\alpha_v$ decreases, the $Y-\omega$ interaction becomes stronger, and there is a suppression of the particles with strange content for high density values. The $\Xi^0$ soon disappears, and the fraction of  $\Xi^-$ becomes less relevant at high densities. It is easier to see the hyperon suppression in a
plot of the strangeness fraction as shown in Fig.~\ref{F3}. 
Exactly the same relation between $\alpha_v$ and strangeness, with obvious 
consequences on the stiffness of the EoS was found in \cite{Weiss2}, an expected
behaviour since our choice of constants was primordially based on this reference.
We also note that although  $U_\Xi$ becomes less repulsive with the decrease of $\alpha_v$, less hyperons tend to be created. The reason is that both the repulsive vector and the attractive scalar channels increase with the decrease of $\alpha_v$. When $\alpha_v$ is small, the strong attractive channel dominates for low densities, but the repulsive one dominates at high densities. This creates a weakly repulsive channel at nuclear saturation point, and the hyperon  suppression for densities much above this density. We also note that for $\alpha_v = 0.0$ the code stops converging at not too high densities. This fact has already been discussed in the literature~\cite{Rafa,Magno,Myung} and happens because the nucleon effective mass becomes zero.
We plot the fraction of particles for the GM1LM in Fig.~\ref{F2} for different values of  $\alpha_v$.  

\begin{figure*}[ht]
\begin{tabular}{cc}
\includegraphics[width=5.6cm,height=6.2cm,angle=270]{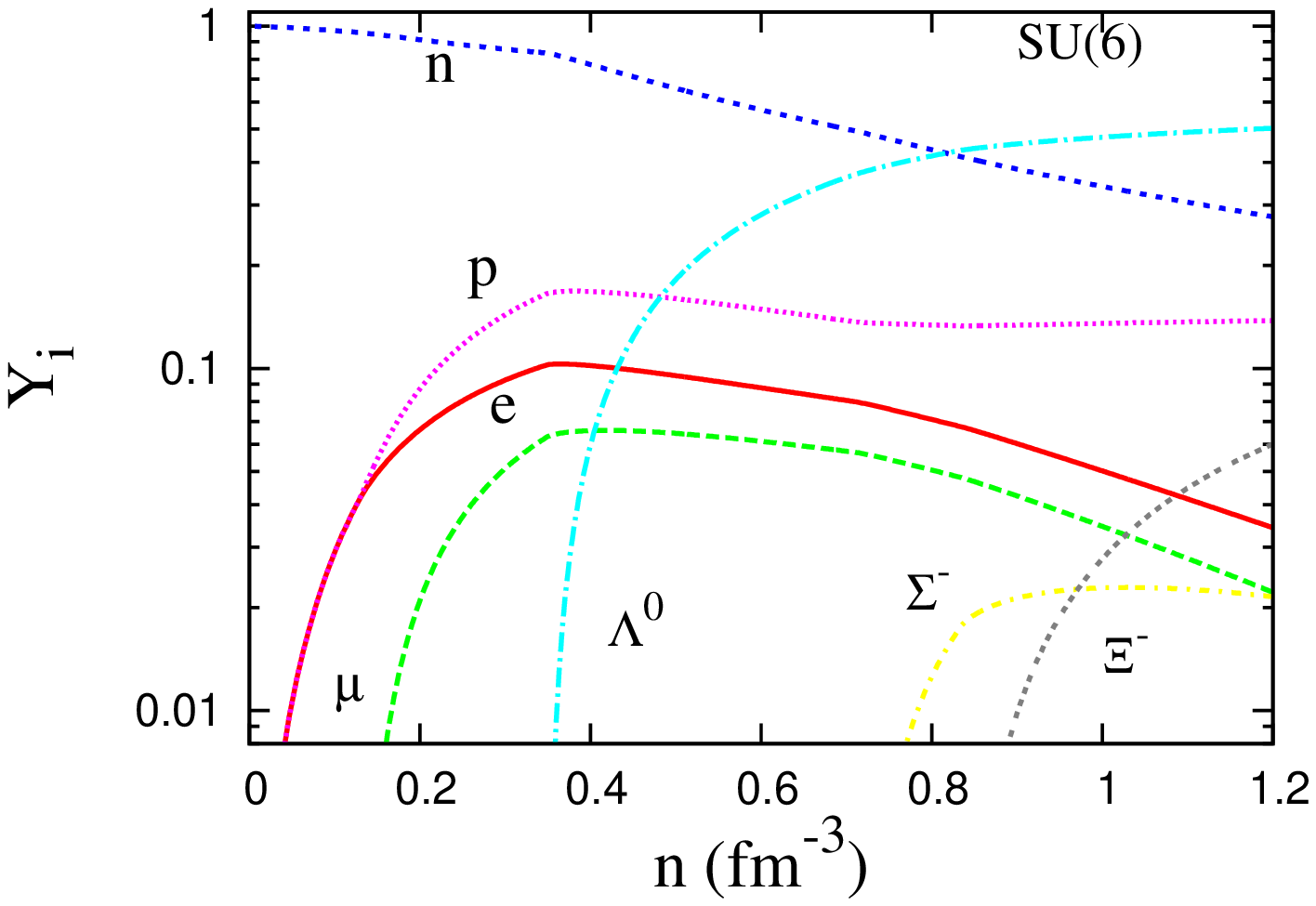} &
\includegraphics[width=5.6cm,height=6.2cm,angle=270]{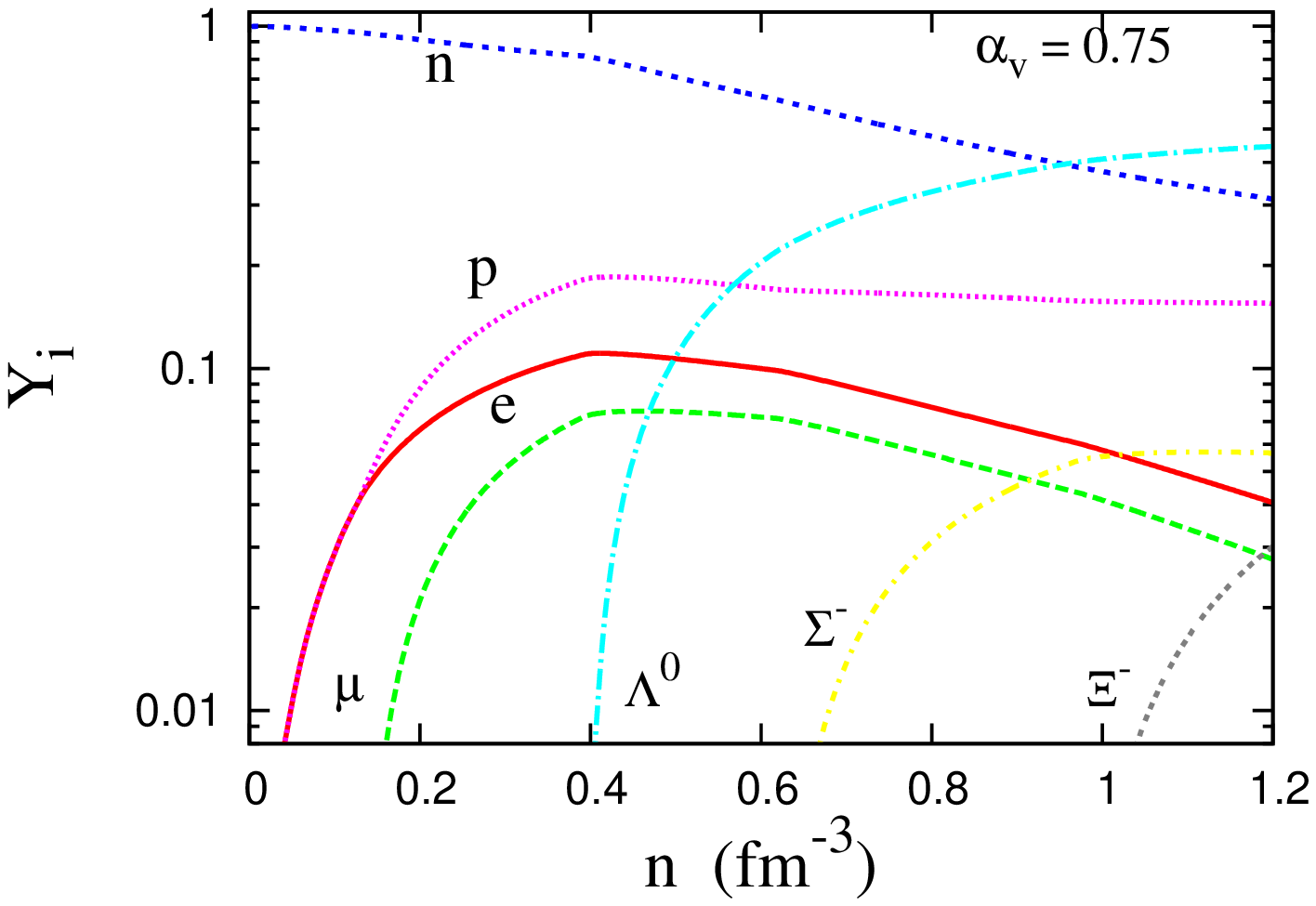} \\
\includegraphics[width=5.6cm,height=6.2cm,angle=270]{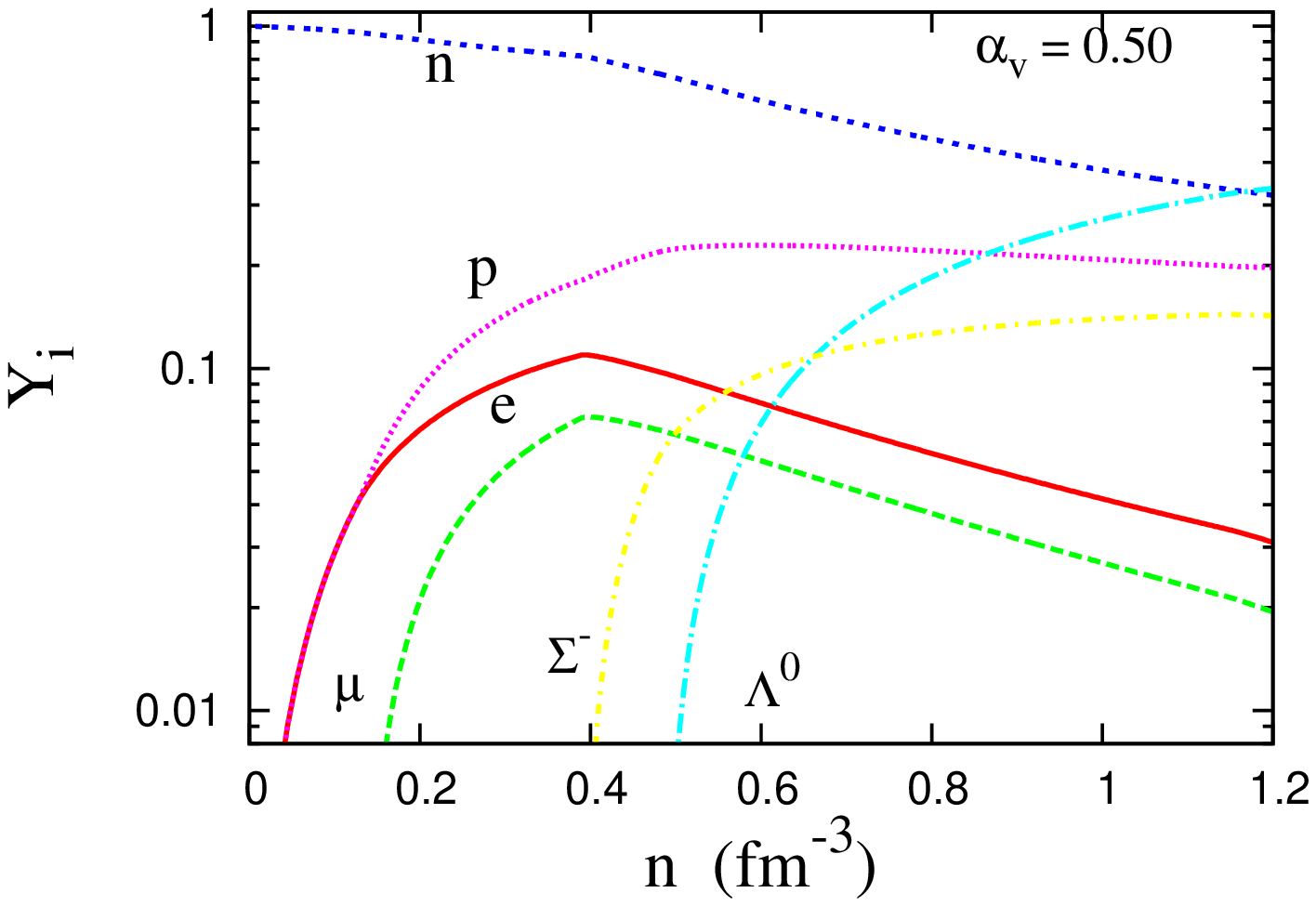} &
\includegraphics[width=5.6cm,height=6.2cm,angle=270]{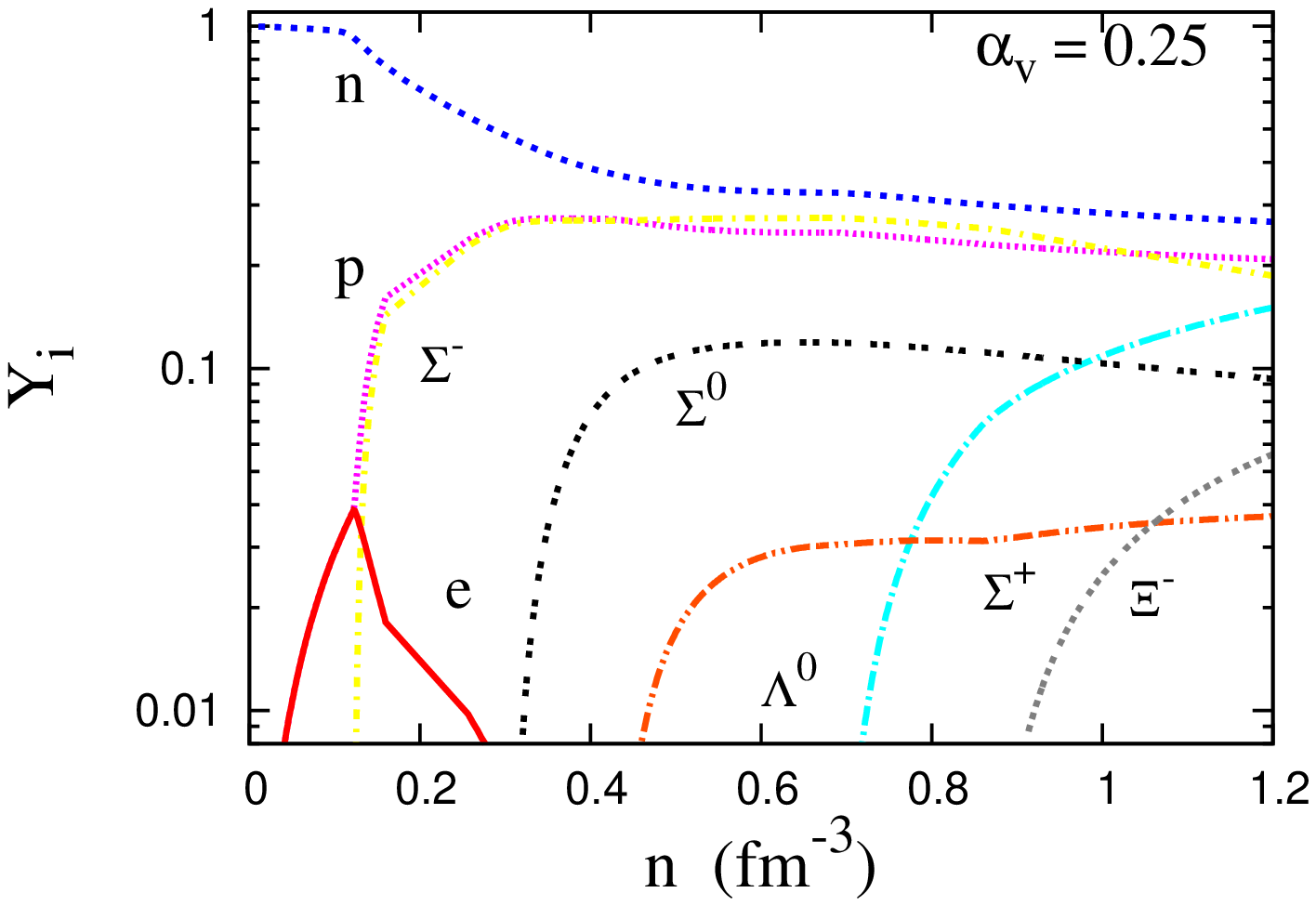} \\
\end{tabular}
\caption{Fraction of particles with the GM1LM parametrization for some different values of  $\alpha_v$. } \label{F2}
\end{figure*}

 The main difference between the GM1 and the GM1LM as far as the fraction of particles is concerned refers to 
the arising of the $\Sigma^-$,  which appears with the GM1LM and is absent with the GM1, except when we use the GC parametrization. The reason is that the $\phi$
 meson couples stronger to the $\Xi$ than to the $\Sigma$. Also, as an additional source of repulsion, the $\phi$ meson  increase the suppression of hyperons
 at high densities and  easing the production of different particles. With the decrease of $\alpha_v$, the $\Sigma$ density onset is lowered.
 The reason is that the $\Sigma -\phi$ coupling becomes  weaker and weaker, while the $\Lambda-\phi$ coupling becomes  stronger. For $\alpha_v = 0.50$,
 the $\Sigma^-$ becomes the first hyperon to appear, while for $\alpha_v = 0.25$, the repulsive channel is so weak for the $\Sigma$ that, alongside the strong attractive channel, the  $\Sigma^-$ arises at the density of 0.13 $fm^{-3}$, even below the nuclear saturation density. This is the effect of the  very low $U_\Sigma$ = -157 MeV, which contradicts all physical expectations.
As pointed out earlier, the  hyperon suppression at hight densities with the decrease of $\alpha_v$, and also with the inclusion of the $\phi$ meson can be best seen in a plot of the  strangeness fraction $f_s$ instead of the individual fraction of particles. $f_s$ is normally defined as:

\begin{equation}
f_s = \frac{1}{3}\frac{\sum_i n_i |s_i|}{n} , \label{s26}
\end{equation}
where $s_i$ is the strangeness of the $ith$ baryon.  

For a matter of clarity, so that the curves do not overlap, we have decided to plot the curves related to the
GC parametrization alongside those of GM1LM, although the $\phi$ meson is not present in the GC set and is always present in LM.

\begin{figure*}[ht]
\begin{tabular}{cc}
\includegraphics[width=5.6cm,height=6.2cm,angle=270]{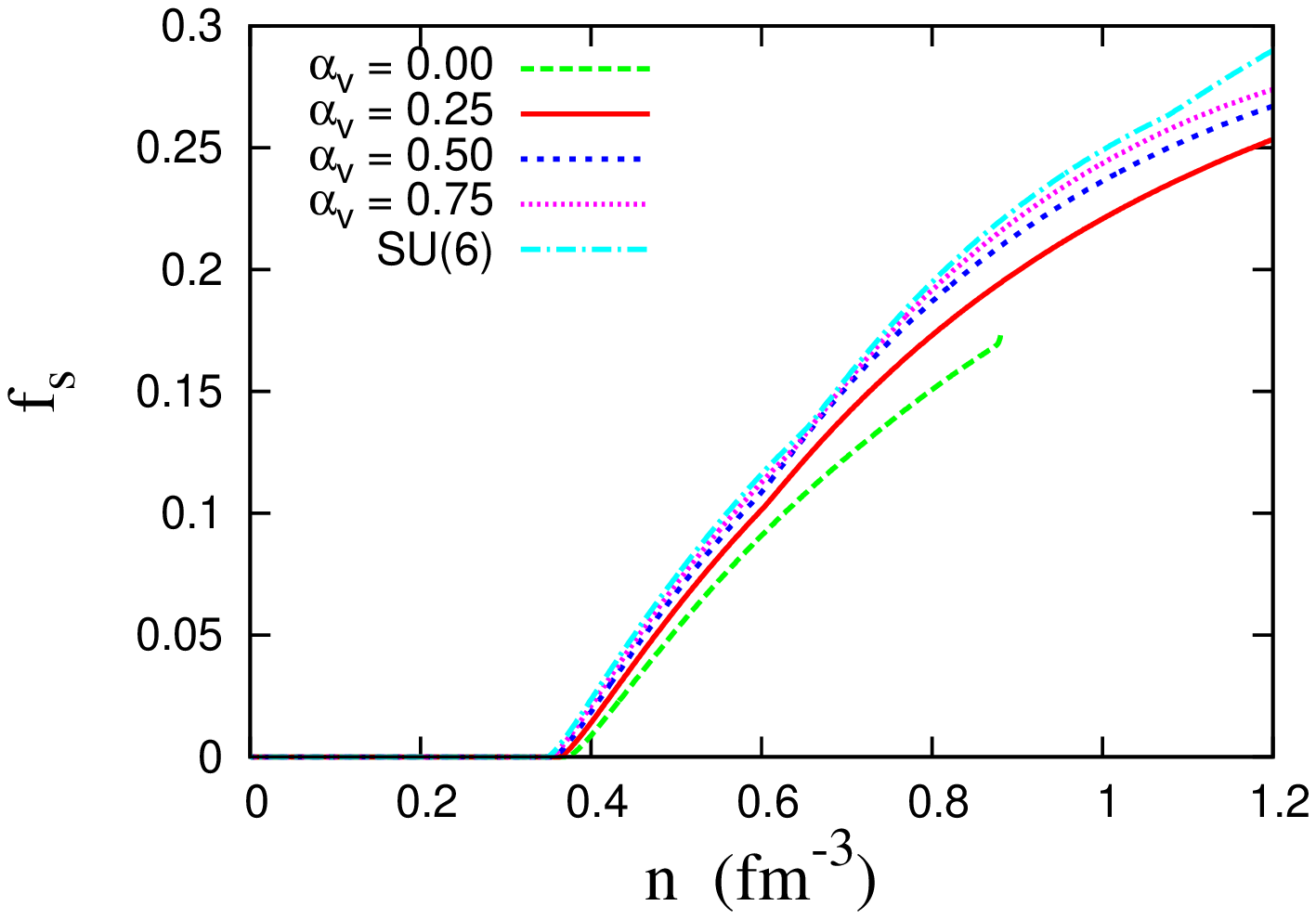} &
\includegraphics[width=5.6cm,height=6.2cm,angle=270]{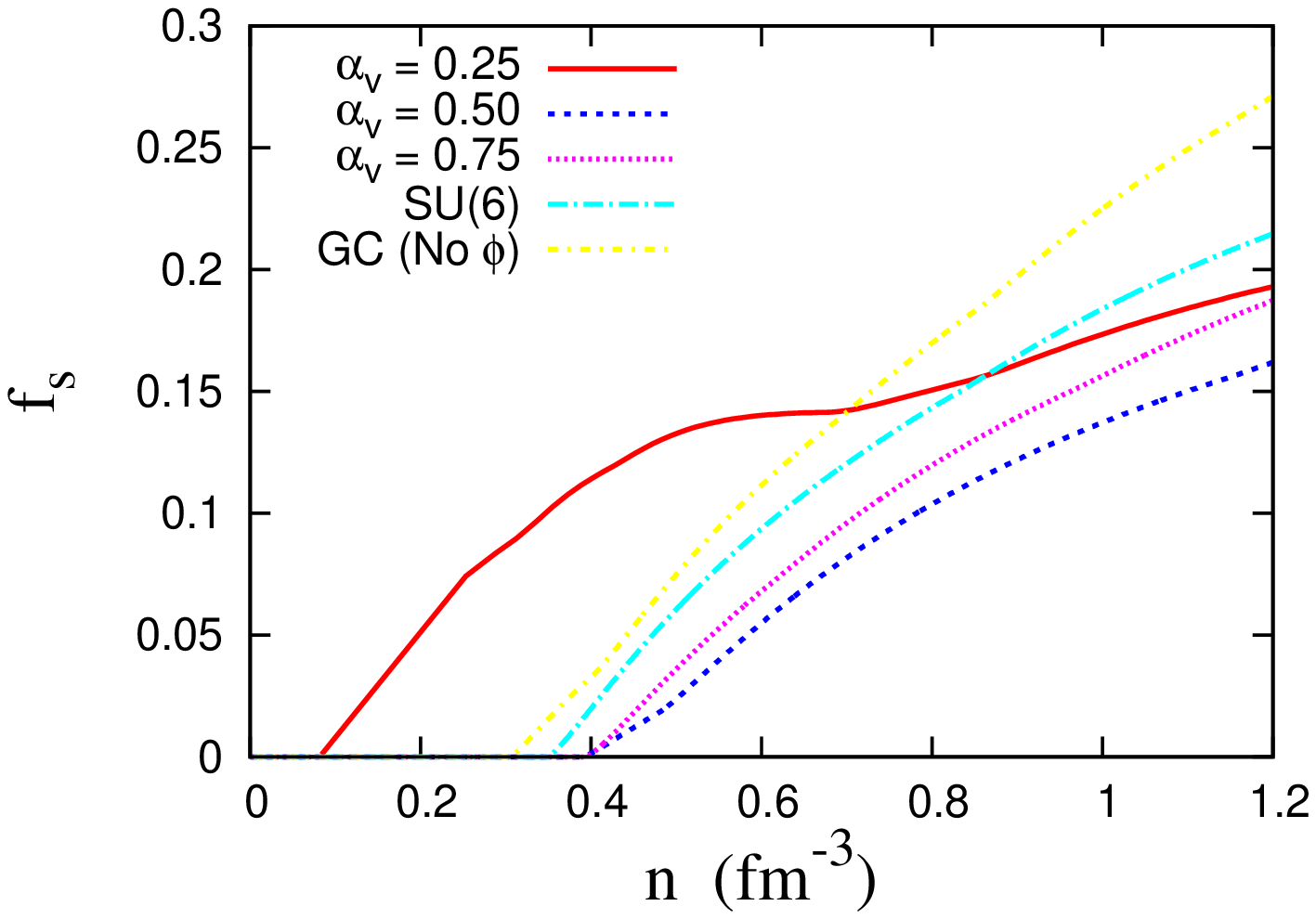} \\
\end{tabular}
\caption{Strangeness fraction for several values of $\alpha_v$ with the GM1 (left) and GM1LM (right) parametrizations. We also include GC parametrization for comparison.}
\label{F3}
\end{figure*}

Wee see that there is  a connection between $\alpha_v$ and $f_s$. A lower $\alpha_v$ results in a  lower $f_s$ both with  GM1 and with the  GM1LM parametrizations
 (except when $\alpha_v=0.25$, GM1LM), due to the strong hyperon suppression at hight densities as already discussed. Moreover, GM1LM produces a  lower $f_s$ when compared with the GM1  due to the new vector channel that increases the chemical potential for the hyperons, hindering their production at high densities. In GM1LM when $\alpha_v = 0.25$, the onset of $\Sigma^-$ at very low densities makes the  strangeness fraction  nonzero even below the saturation density. Due to this effect, from this point on, we no longer use this parametrization.

One of the most important quantities of nuclear and hypernuclear matter is the equation of state (EoS). We plot the EoS for the parametrizations discussed above in Fig.~\ref{F4}.

\begin{figure*}[ht]
\begin{tabular}{cc}
\includegraphics[width=5.6cm,height=6.2cm,angle=270]{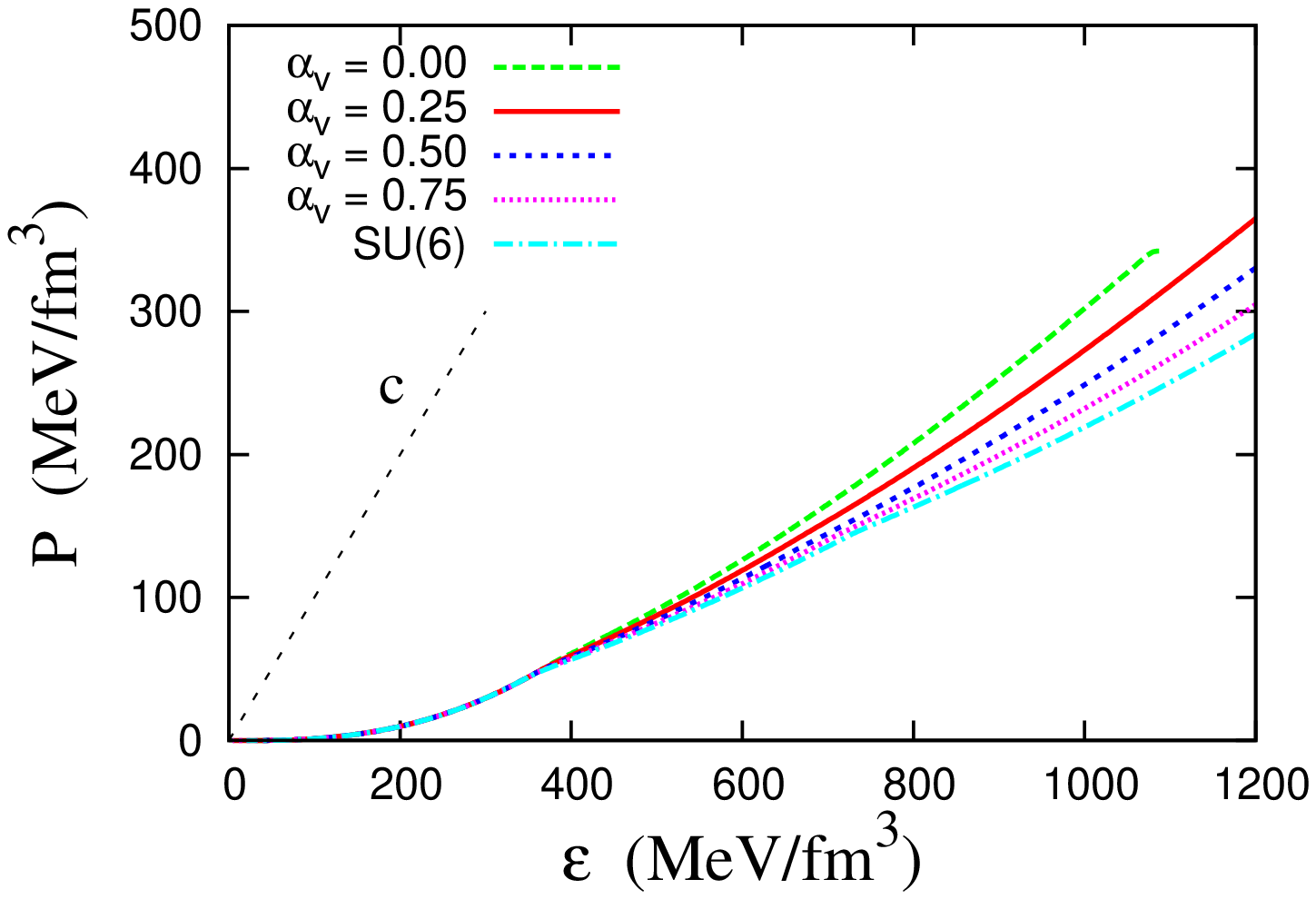} &
\includegraphics[width=5.6cm,height=6.2cm,angle=270]{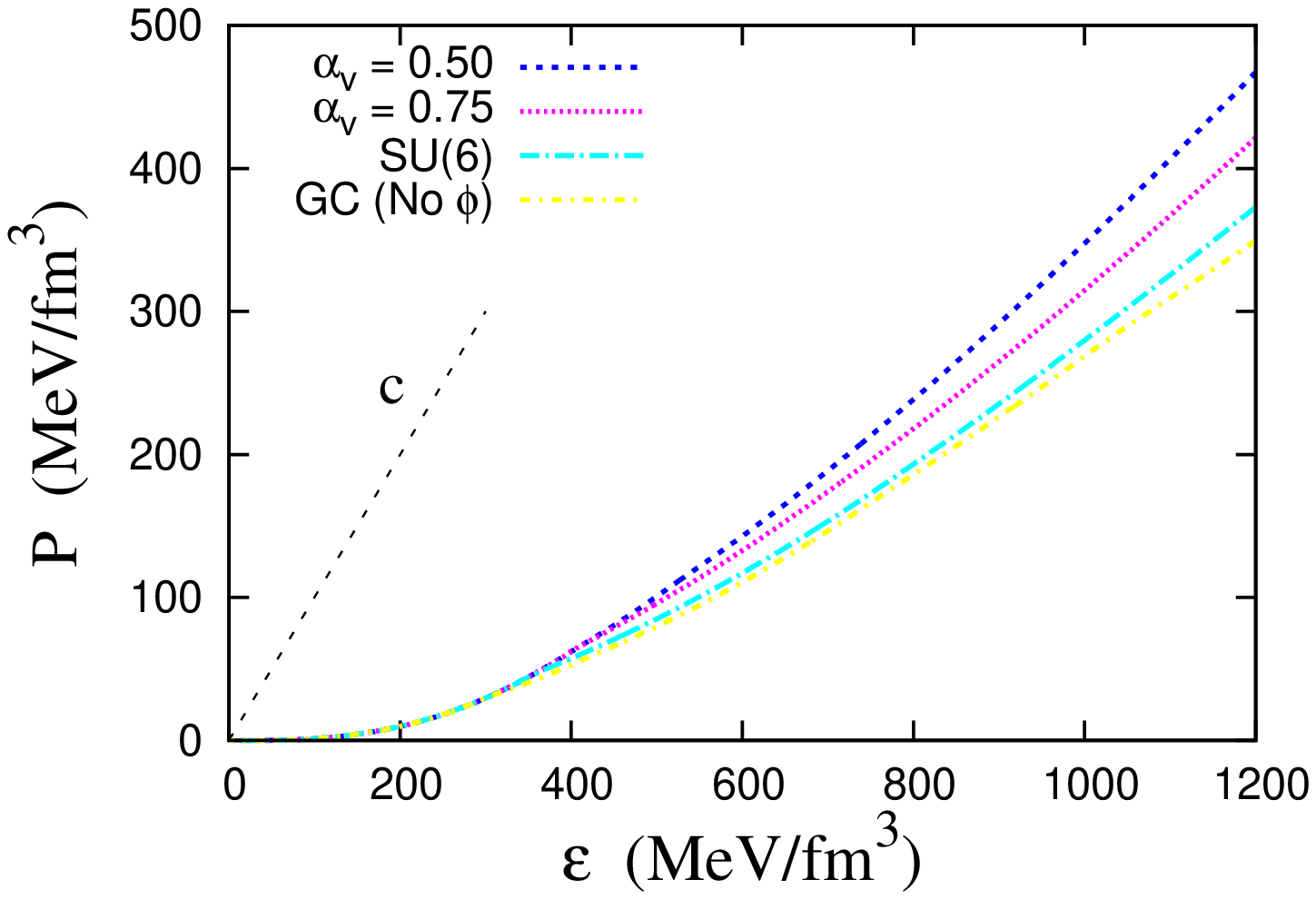} \\
\end{tabular}
\caption{EoS for GM1 (left) and GM1LM (right) for GC and SU(3) group with several values of $\alpha_v$. The `c' is the causality limit.} \label{F4}
\end{figure*}

The relation between $\alpha_v$ and $f_s$ can also be observed in the EoS. Lower values of $\alpha_v$ produce stiffer EoS. This was expected, since lower $f_s$ indicates less hyperons, and the effect of softening of the EoS caused by the hyperons is well known~\cite{Glen,Haensel,Glen3,Rufa}. It is interesting to note that the stiffer EoS is obtained with less repulsive hyperon potentials. This indicates that the influence of the hyperon potential depths as pointed in ref.~\cite{Weiss} is only secondary. The ruling term for the stiffness of the  EoS is the strength of the $Y-\omega$ interaction. In GM1LM we have alongside the $Y-\omega$ interaction, the $Y-\phi$ one. These combined effects produce very stiff EoS as we can see on the right side of Fig.~\ref{F4},  even for low repulsive hyperon potentials.
 The {\bf c} line plotted refers to the causality limit. Since our results are derived from a relativistic model, this constraint is never violated.

We have also analyzed the speed of sound $v_s$,  defined as:
\begin{equation}
v_s = \sqrt{ \bigg | \frac{\partial p}{\partial \epsilon} \bigg |}, \label{c6e3}
\end{equation}
since it is an important quantity in the study of phase transitions. Although there is no experimental measurement of the speed of sound at high densities, results derived from QCD impose a limit of  $v_s = 1/\sqrt{3}$ $\simeq$ 0.58 (c = 1) for the quark-gluon plasma~\cite{Hohler,Hostler,Chamel}. Methods to measure the speed of sound are proposed in modern literature~\cite{Renk,Neufeld} through Mach cones. Since the density of a possible
hadron-quark phase transition is not known, a measure of $v_s > 0.58$ excludes the possibility of quark matter at  the densities where such speed is obtained. We plot the speed of sound in Fig.~\ref{F5}

\begin{figure*}[ht]
\begin{tabular}{cc}
\includegraphics[width=5.6cm,height=6.2cm,angle=270]{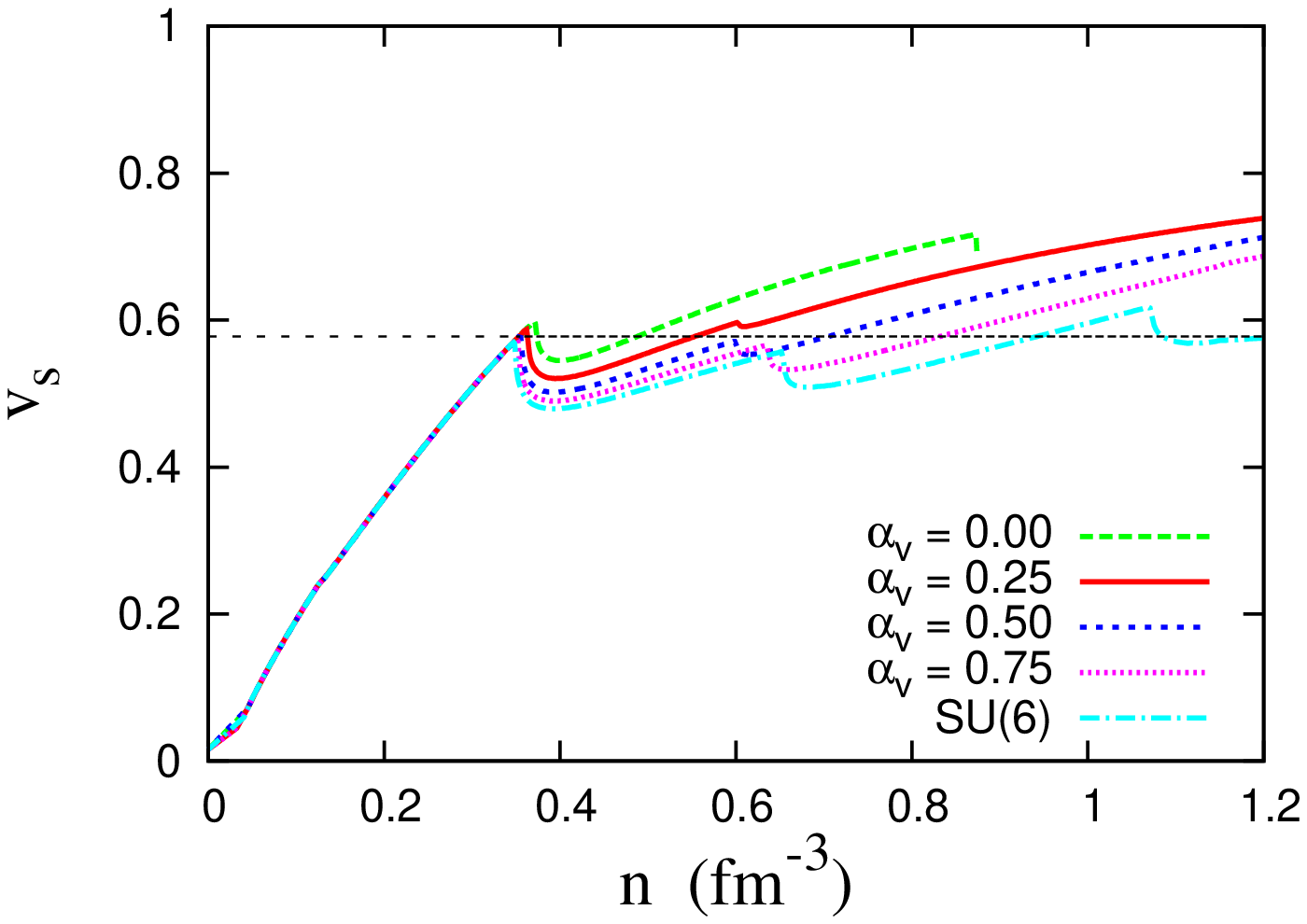} &
\includegraphics[width=5.6cm,height=6.2cm,angle=270]{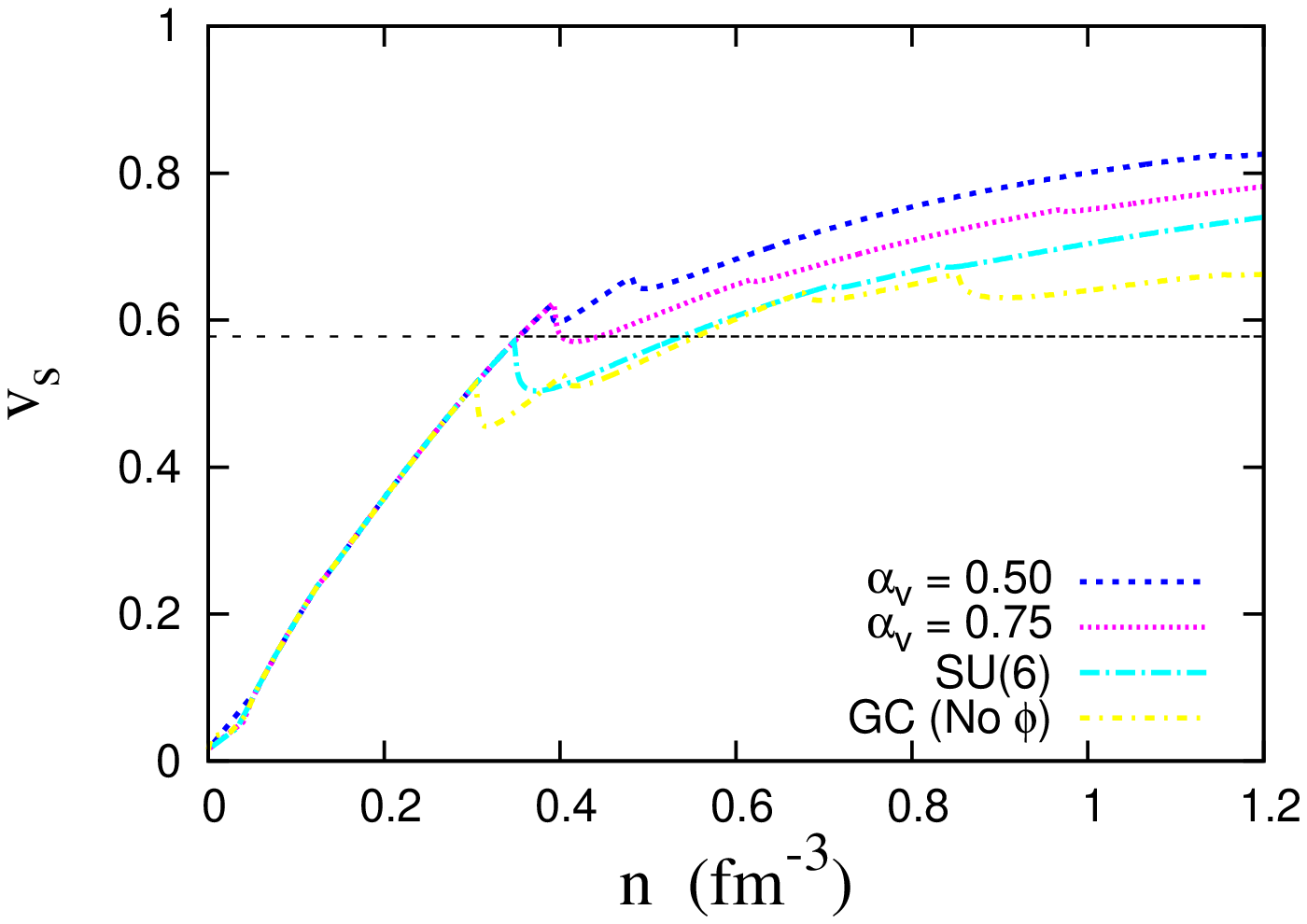} \\
\end{tabular}
\caption{Speed of sound in dense hypernuclear matter for GM1 (left) and GM1LM (right). The 
horizontal line is the QCD limit of $v_s$ for quark matter.} \label{F5}
\end{figure*}

As expected, the speed of sound is the same for all parametrizations while hyperons are not present. The onset of hyperons reduces the speed of sound
due to the softening of the related EoS,  resulting in 
a connection between $\alpha_v$ and $v_s$.
A stiffer EoS, obtained with lower values of $\alpha_v$, yields a 
higher value for the speed of sound at high densities.

Without hyperons the breaking of the QCD theoretical limit for the speed of sound occurs at $0.352 ~fm^{-3}$. The main question therefore is in which parametrizations
 the first hyperon arises below this density. When the first hyperon arises 
before the density of $0.352 ~fm^{-3}$, the density at which the speed of sound
exceeds $0.58$ increases substantially.  We display in Table \ref{T6} the density in which the speed of sound surpasses the QCD speed limit for different
 values of $\alpha_v$.

  \begin{table}[ht]
\begin{tabular}{|c|c|c||c|c|c|}
\hline
 Model & $\alpha_v$ & $v_s > 0.58$ at: &  Model & $\alpha_v$ & $v_s > 0.58$ at:   \\
\hline
 GM1 & SU(6)& 6.16 $n_0$  & GM1LM & SU(6) & 3.51 $n_0$    \\
\hline
 GM1 &  0.75 & 5.44 $n_0$ & GM1LM& 0.75& 2.30 $n_0$   \\
 \hline
GM1 & 0.50 & 4.60 $n_0$ & GM1LM & 0.50 & 2.30 $n_0$   \\
\hline
GM1 & 0.25 & 2.30 $n_0$ & &  &    \\
\hline
GM1 & 0.0 & 2.30 $n_0$  & GM1 & GC & 3.62 $n_0$    \\
\hline
\end{tabular}
 \caption{Number density in which the speed of sound exceeds the theoretical limit of QCD for GM1 and GM1LM with different parametrization sets.}\label{T6}
 \end{table}
 
 We see that for the sets GC, SU(6), $\alpha_v = 0.75$ and $\alpha_v = 0.50$ 
with the GM1 parametrization and the SU(6) with the GM1LM, the first hyperon 
appears  below $0.352~fm^{-3}$, resulting in the fact that the QCD limit of the 
speed of sound is attained at higher density values.
Later we see that this fact  has implications when we analyze some experimental 
constraints.

\subsection{Constraints}

 In order to validate our proposal, we confront it with some experimental values
and astrophysical observations.
 The first experimental constraint (which we call here  EC1) is the asymmetric  coefficient $S_0$ and the symmetry energy slope $L$ as defined in ref.~\cite{Tsang}:

 \begin{equation}
 S(n) = S_0 - L\epsilon + \frac{1}{2}K_{sym} \epsilon^2 + O(\epsilon^3), \label{c6e5}
 \end{equation}
  \begin{equation}
 \quad \epsilon = \frac{n_0 - n}{3n_0}, \quad \mbox{and} \quad L = 3n_0 \frac{dS}{dn} \bigg |_{n_0} . \label{c7}
 \end{equation}

 The values of $S_0$ are 32.5 MeV for GM1 and GM3, and 37.4 MeV for NL3, while the slope $L$ assumes 94 MeV for the GM1, 90 MeV for the GM3 and 118 MeV for the NL3~\cite{Rafa}. We analyze these tree parametrizations in the light of two experimental data described in ref.~\cite{Tsang}, obtained from heavy ions collision (HIC, hatched area in green) and isobaric analog states (IAS, hatched area in pink). The experimental values of $S_0$ and $L$ and the theoretical previsions obtained with GM1, GM3 and NL3 are presented in Fig.~\ref{F6}. 

\begin{figure}[ht] 
\begin{centering}
 \includegraphics[angle=270,
width=0.4\textwidth]{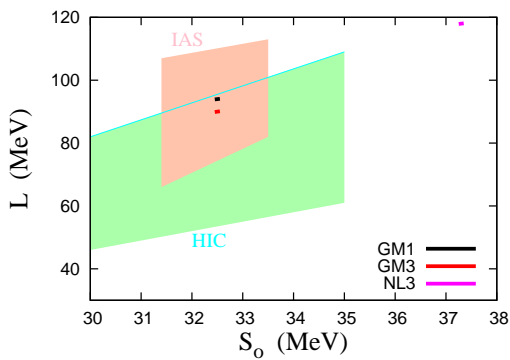}
\caption{EC1:  $S_0$ and $L$ experimental values obtained for HIC and IAS, and the prevision of GM1, GM3 and NL3 parametrizations.} \label{F6}
\end{centering}
\end{figure}
We see that while GM1 and GM3  agree with both experimental measurements, the NL3 parametrization is completely out of the allowed region. Since we expect that neutron stars are  neutron rich systems, the symmetry energy plays a crucial role in the description of these objects. Being the NL3 parametrization in disagreement with the experimental observation, this parametrization should be avoided at least in the description of neutron star properties.

The second experimental constraint (EC2) is the nucleon potential in symmetric matter. From measurements of kaon production in HIC performed with the kaon spectrometer (KaoS)~\cite{JSB}, an upper limit was established till at least two times the nuclear saturation density. In QHD based models the nucleon potential in symmetric matter is defined as:
\begin{equation}
U_N =  g_{NN\omega}\omega_0 - g_{NN\sigma}\sigma_0 . \label{s28}
\end{equation}

The third experimental constraint (EC3) is  the  pressure of symmetric nuclear matter up to almost five times the nuclear density as obtained in ref~\cite{Daniel}. The previsions obtained with GM1, GM3 and NL3 and the experimental upper limit of the nucleon potential withdrawn from ref.~\cite{JSB} are plotted in Fig.~\ref{F7} left, and the pressure (hatched area) withdrawn from ref.~\cite{Daniel} and the previsions are plotted in Fig.~\ref{F7} right.

\begin{figure*}[ht]
 \begin{center}
\begin{tabular}{cc}
\includegraphics[width=5.6cm,height=6.2cm,angle=270]{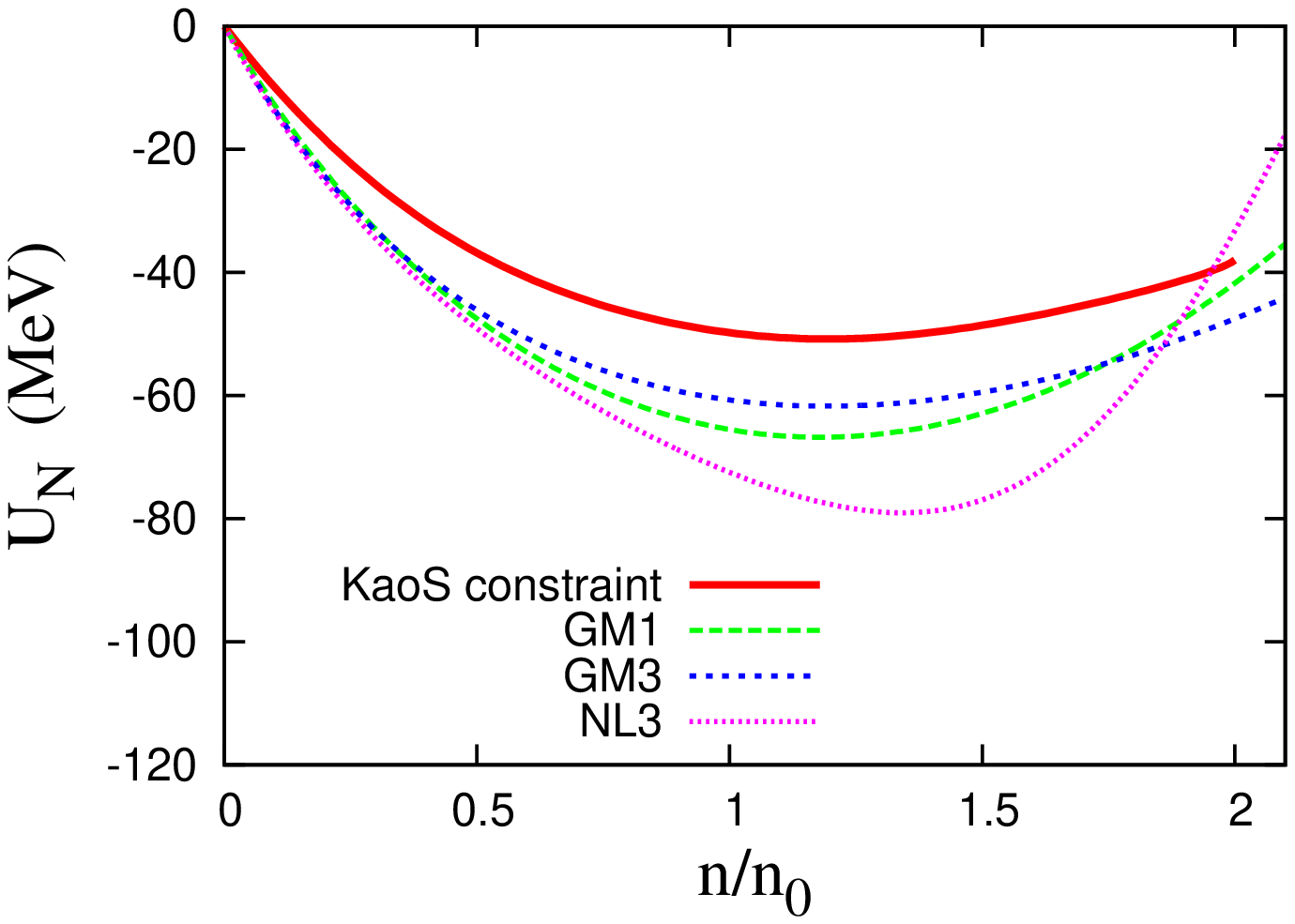} &
\includegraphics[width=5.6cm,height=6.2cm,angle=270]{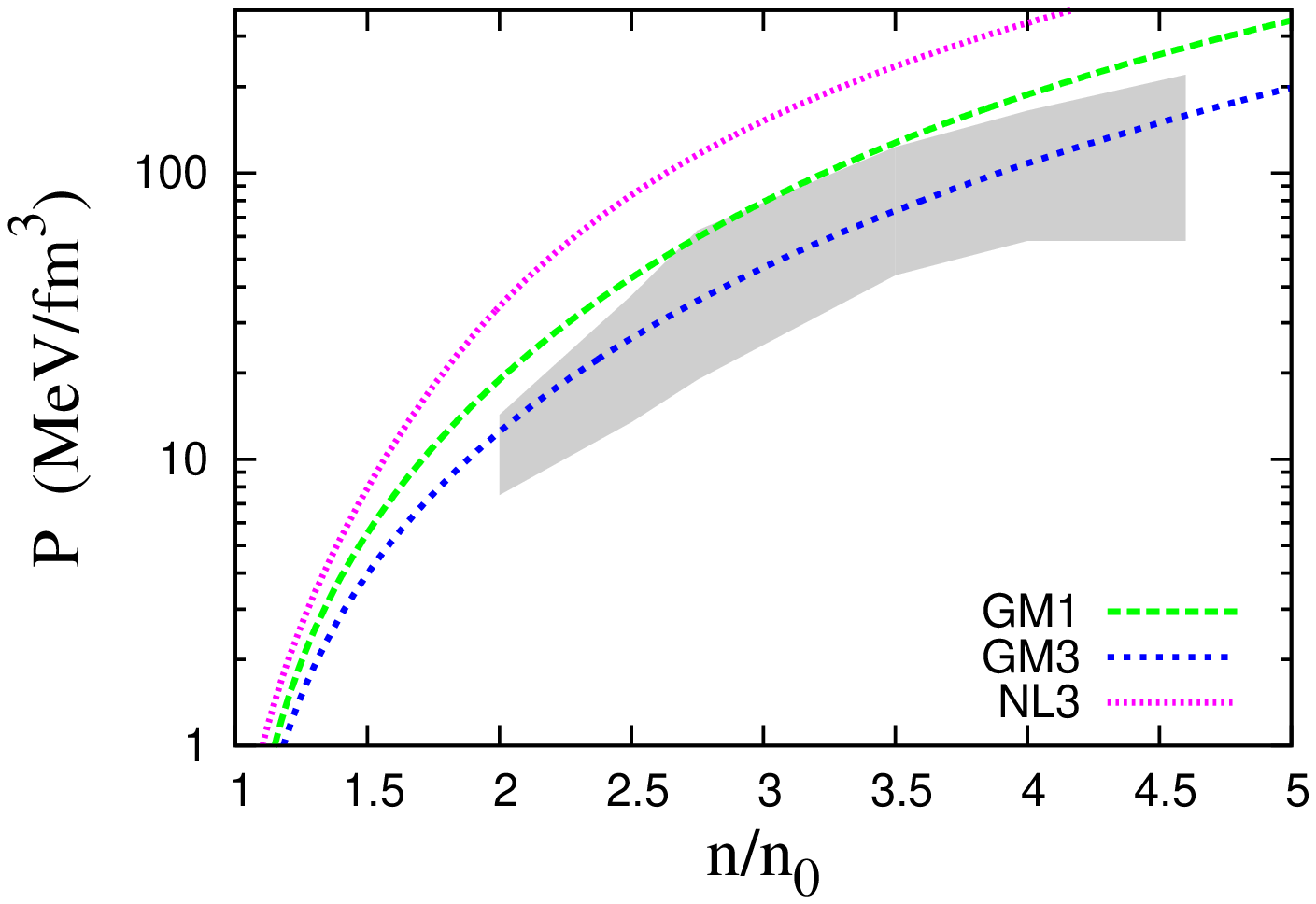} \\
\end{tabular}
\caption{EC2 (left): experimental value of  $U_N$ from kaon production up to  two times the saturation point ($n_0 = 0.153~fm^{-3}$).
  EC3 (right) Experimental determination of the pressure in symmetric nuclear matter.} \label{F7}
\end{center}
\end{figure*}

Both  constraints point towards the same direction: the EoS must be soft for densities not much above the nuclear saturation point. 
Wee see that NL3 fails again when confronted with  the EC2 and EC3. On the other hand, GM3 fulfills all the experimental constraints we have analyzed.
The GM1 parametrization has a more delicate situation. Whilst GM1 is in accordance with EC1 and EC2, it fails to describe EC3. So, at first glance,
  this parametrization should be be ruled out. However, as pointed in ref.~\cite{Daniel}, the experimental results do not rule out hyperon creation
 nor even more exotic pictures as quark-hadron phase transitions.
 The softening of the EoS due to the hyperon onset could be the key to reconcile experiment and theory.

Note that in symmetric nuclear matter, the chemical potential of the protons is equal to the chemical potential of the neutron.
Now we define the hypernuclear symmetric matter imposing that the hyperon chemical potential be equal to the  chemical potential of the nucleons:
\begin{equation}
 \mu_Y = \mu_p = \mu_n.
\end{equation}
This choice implies that only symmetric nuclear matter exists until the density is high enough so that the creation of strange particles becomes
energetically favorable, softening the EoS. We plot the pressure of hypernuclear symmetric matter alongside the experimental constraint EC3 (hatched area) in Fig.~\ref{F8}.

\begin{figure*}[ht]
  \begin{tabular}{cc}
 \includegraphics[width=5.6cm,height=6.2cm,angle=270]{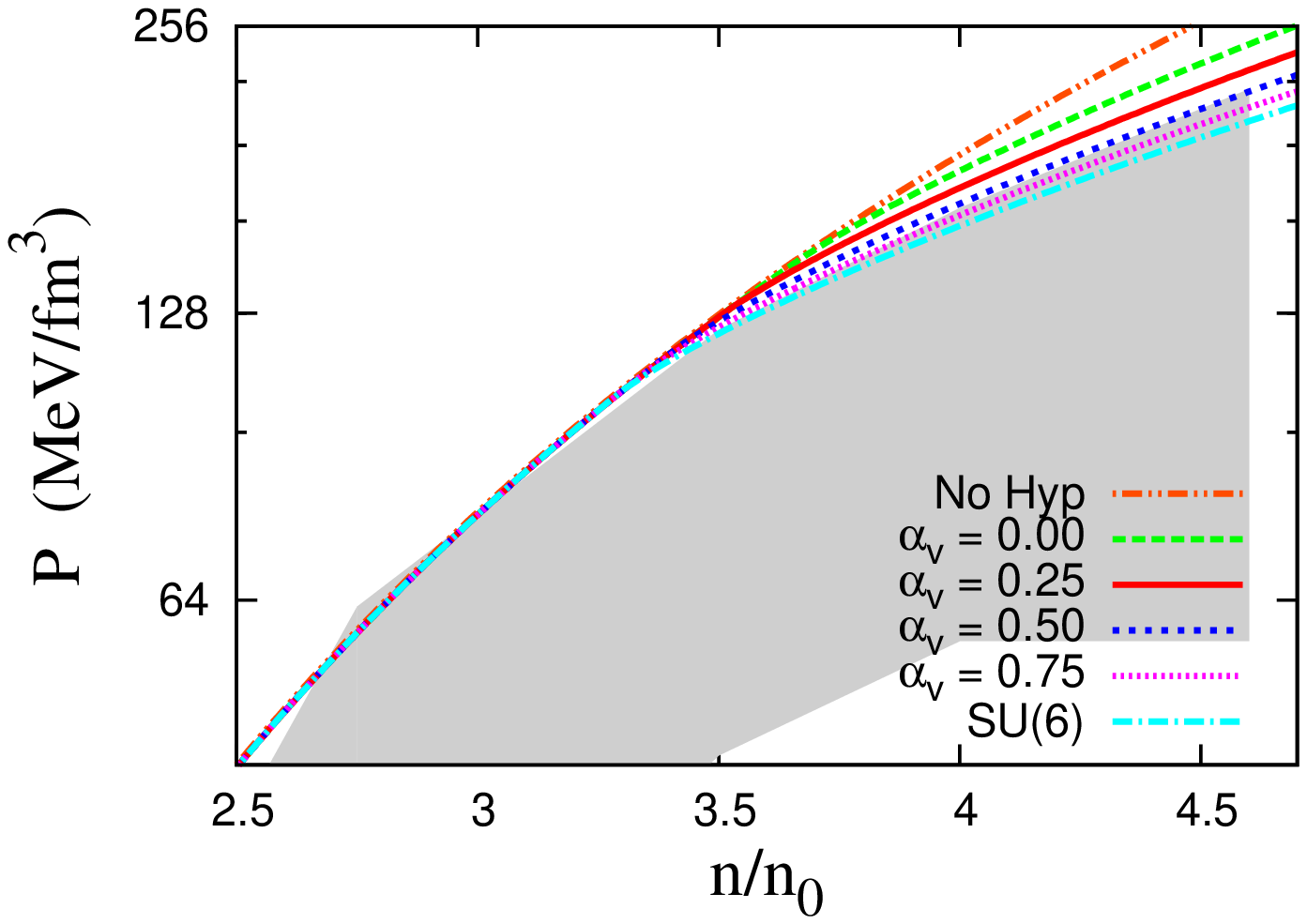} &
\includegraphics[width=5.6cm,height=6.2cm,angle=270]{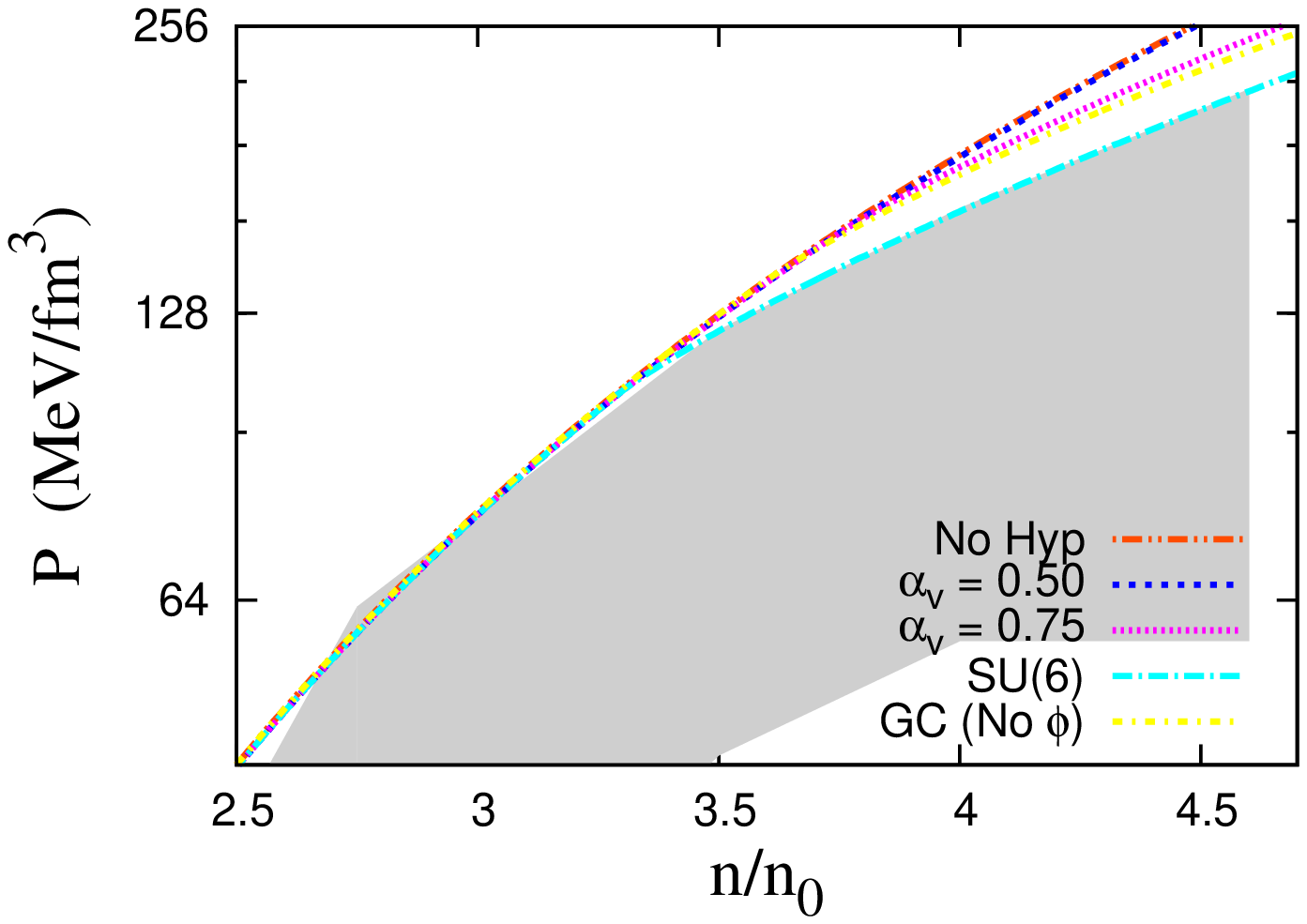} \\
\end{tabular}
\caption{Pressure of hypernuclear symmetric matter in GM1 (left) and GM1LM (right) parametrizations for different values of $\alpha_v$.} \label{F8}
\end{figure*}

We see that for $\alpha_v = 1$, $\alpha_v = 0.75$ and $\alpha_v = 0.50$ with GM1, and the SU(6) parametrization with GM1LM, the prediction of pressure falls
 again in the experimental area at high densities. It is interesting to note that the normally used  GC parametrization fails to reproduce the experimental
 constraint, and therefore, should be taken with care.

 As pointed out earlier, the speed of sound has implications which relate to the EC3.
 With the exception of the GC,
all parametrization sets that agree with the EC3, break QCD theoretical limit 
of the speed of sound for densities above three times the nuclear saturation density.
Moreover, according to our proposal, all parametrization sets that agree 
with the EC3, have the $\Lambda^0$ as the first hyperon that arises, and its onset 
needs to occur at densities below $0.352 fm^{-3}$.

 It is also important to note that although some of the GM1 and GM1LM hypernuclear parametrizations are in agreement with the experimental data for high densities,
 they are  still  in disagreement at low densities.  However, this could be due to the fact that we  ignore meson production,
 important at sub-threshold densities as pointed in ref.~\cite{JSB,Peter}.

Now we turn towards the constraints obtained from astrophysical observations. The main constraints are the recent observations of the two massive pulsars,
 PSR J1614-2230~\cite{Demo} and PSR J0348+0432~\cite{Antoniadis}, which  indicate that the EoS has to be stiff enough to reproduce 2 $M_\odot$ neutron stars.  Therefore, we have two, in principle, contradictory constraints for low and high densities: the experimental ones point to a soft EoS at low densities~\cite{JSB,Daniel}, and the astrophysical observations favor a stiff EoS at high densities~\cite{Demo,Antoniadis}. We next check all  parametrizations discussed so far to see whether they can fulfill these two features. The mass of the PSR J0348+0432 is $2.01\pm0.04$ $M_\odot$, and we use this value as the first astrophysical constraint (AC1).
The second astrophysical constraint (AC2) is the redshift measurements ($z$) of two neutron stars. A redshift of $z = 0.35$ has been obtained from three different transitions of the spectra of  the EXO0748-676~\cite{Cottam}. This redshift corresponds to $M/R=0.15 M_\odot/km$. Another constraint on the mass/radius relation comes from the observation of two absorption features in the source spectrum of the 1E 1207.4-5209 neutron star, with redshift from $z = 0.12$ to $ z = 0.23$, which gives $M/R=0.069 M_\odot/km$ to $M/R=0.115 M_\odot/km$~\cite{Sanwal}.

It is also worth mentioning that recently the existence of a 2.7 solar mass object named PSR J1311-3430~\cite{Romani} was suggested.
  Although such hypermassive pulsar is not confirmed yet, all previous measurements indicate a lower limit of 2.1 $M_\odot$. We use this value as a possible constraint.

 To check which  EoS is hard enough to reproduce the well known PSR J0348+0432 and also agrees with the redshift measurements,
 we solve the Tolman-Oppenheimer-Volkoff equations~\cite{TOV}, which are the differential equations for the structure of a static, spherically symmetric,
 relativistic star in hydrostatic equilibrium. Also, we use the BPS~\cite{BPS} EoS for the very low density regime to simulate the neutron star crust. We plot the results for GM1 and GM1LM for several values of $\alpha_v$, and the GC parametrization in Fig.~\ref{F9}.
  The properties of the maximum mass of each parametrization are displayed in Table~\ref{T7}.

 \begin{figure*}[ht]
 \begin{tabular}{cc}
\includegraphics[width=5.6cm,height=6.2cm,angle=270]{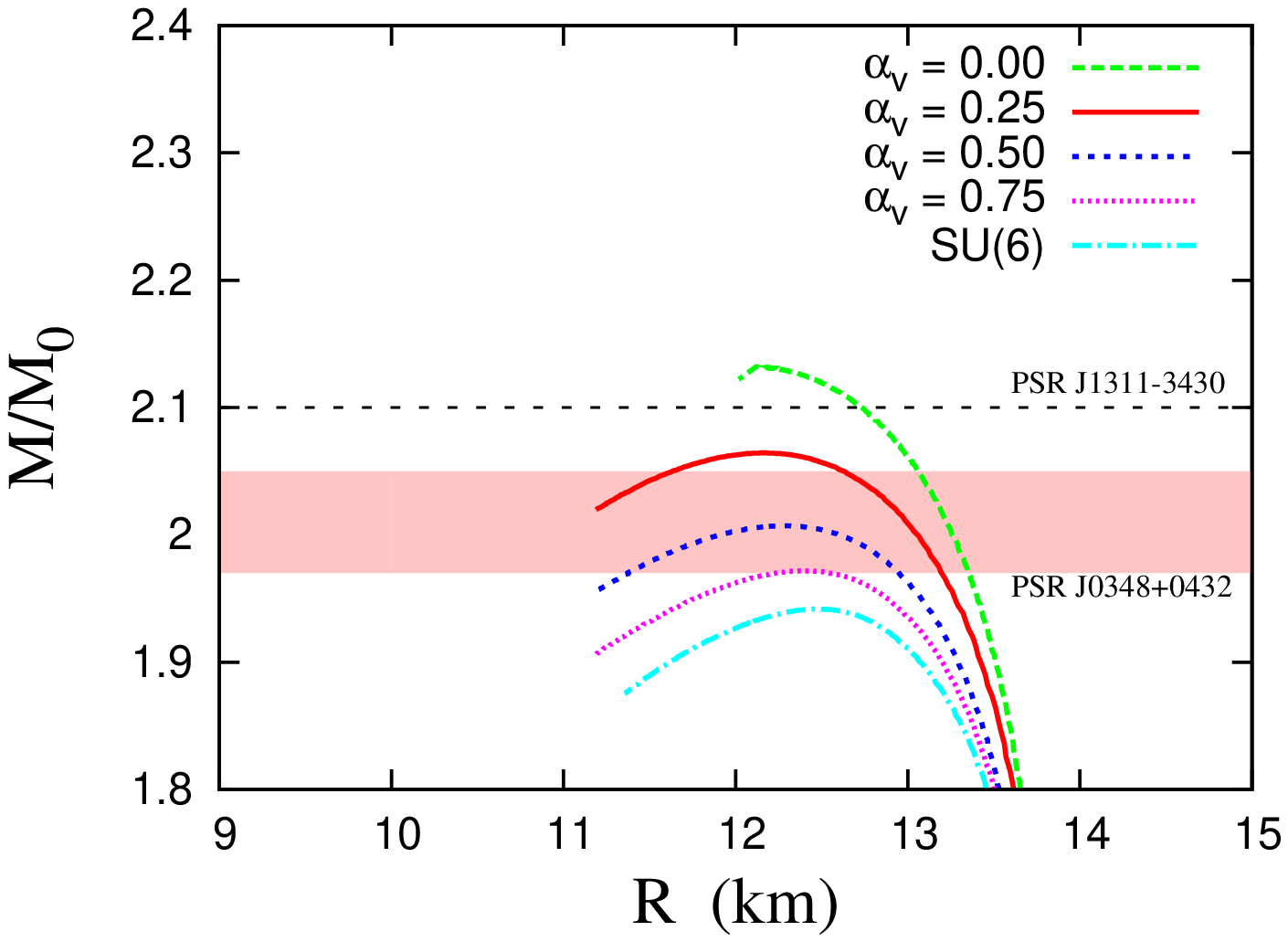} &
\includegraphics[width=5.6cm,height=6.2cm,angle=270]{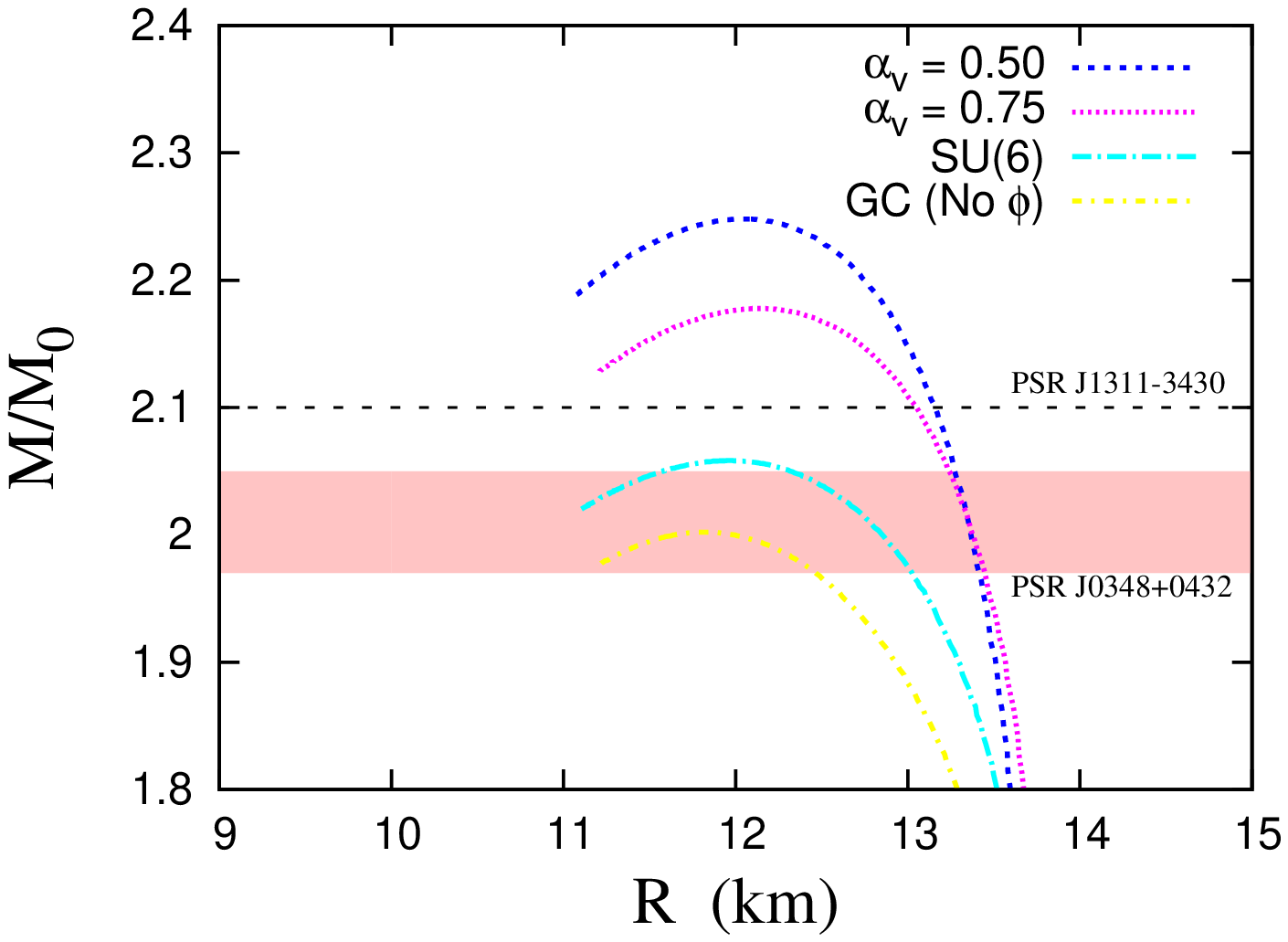} \\
\includegraphics[width=5.6cm,height=6.2cm,angle=270]{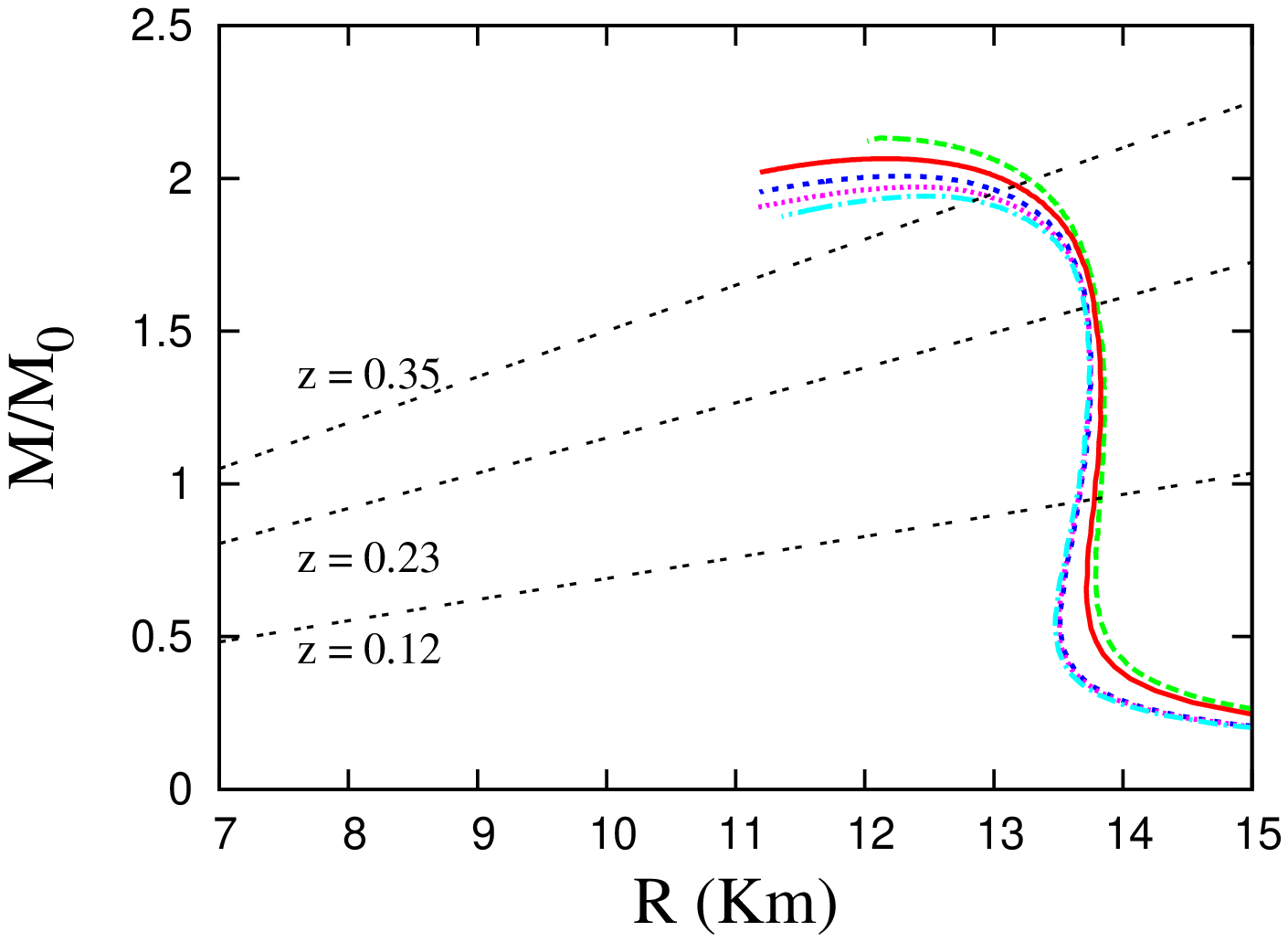} &
\includegraphics[width=5.6cm,height=6.2cm,angle=270]{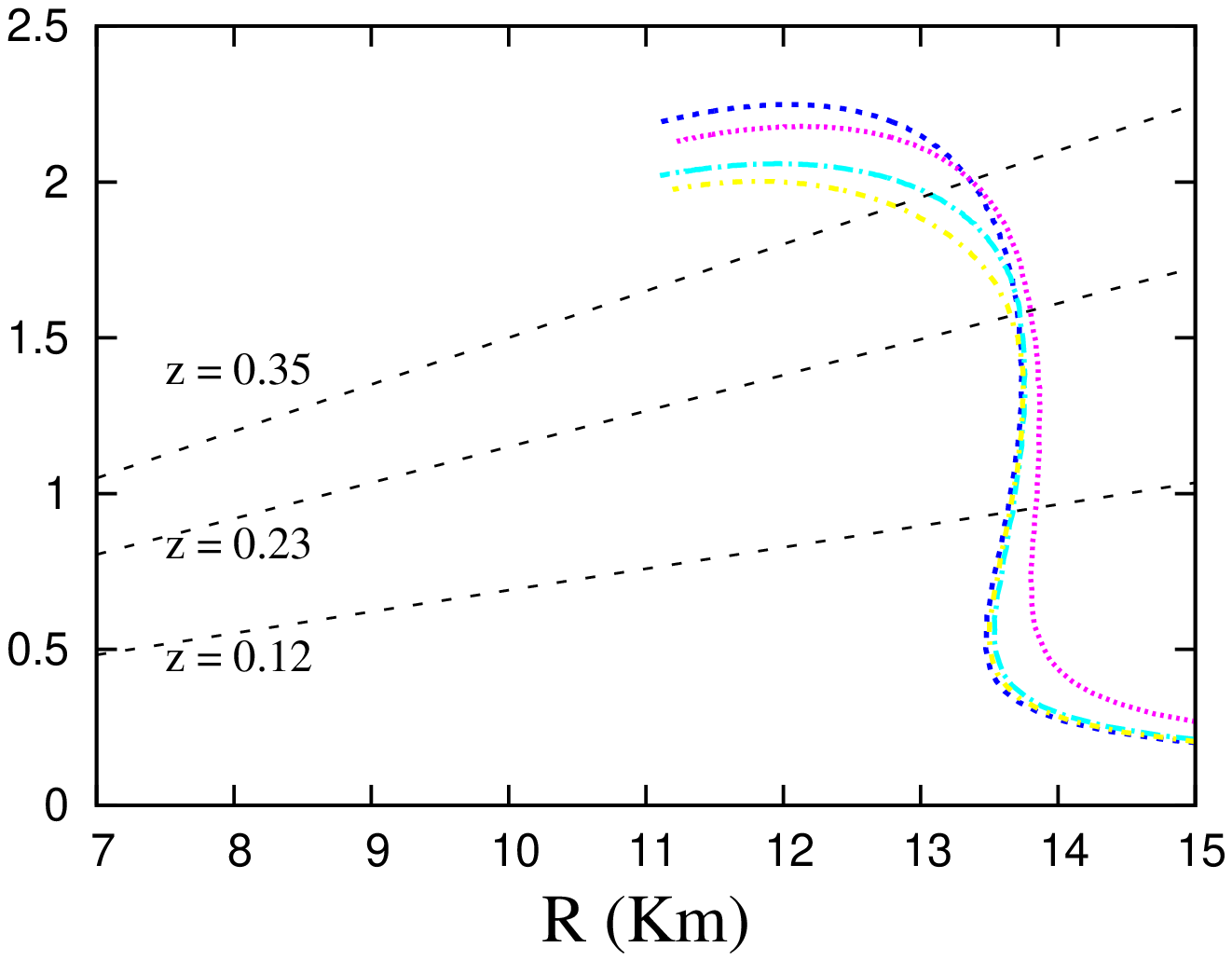} \\
\end{tabular}
\caption{Mass/radii ratio for  GM1 (left) e GM1LM (right) emphasizing the AC1 (above) and the AC2 (below).}\label{F9}
\end{figure*}

  \begin{table}[ht]
\begin{tabular}{|c|c|c|c|c|c|}
\hline
 Model & $\alpha_v$ & $M/M_\odot$ &  R (Km) & $n_c$ $(fm^{-3})$ & $f_s$ at $n_c$.   \\
\hline
GM1 & No hyp. & 2.39   & 11.99 & 0.840 & 0.00  \\
\hline
 GM1 & SU(6)& 1.94   & 12.48 & 0.836 &  0.206    \\
\hline
 GM1 & 0.75 & 1.97  & 12.40 & 0.848 & 0.206 \\
 \hline
GM1 & 0.50 & 2.00  &  12.28 & 0.867 & 0.206  \\
\hline
GM1 & 0.25 & 2.06  & 12.16 & 0.885 & 0.196  \\
\hline
GM1 &  0.0 & 2.13   & 12.09 & 0.891 & 0.178     \\
\hline
GM1 & GC & 2.01   & 11.86 & 0.952 &  0.231    \\
\hline
GM1LM & SU(6) & 2.06   & 11.96 & 0.915 & 0.168    \\
\hline
GM1LM & 0.75 & 2.17   & 12.13 & 0.875 & 0.135   \\
\hline
GM1LM & 0.50 & 2.25   & 12.04 & 0.870 &  0.117    \\
\hline
\end{tabular}
 \caption{Stellar properties obtained with GM1 and GM1LM parametrizations with GC and different values of $\alpha_v$.}\label{T7}
 \end{table}

From Fig.~\ref{F9} and table \ref{T7}, one can see that the only parametrization that fails to describe the AC1 is GM1 with the SU(6) choice of couplings. All  other parametrizations both in GM1 and in GM1LM  yield a maximum mass of at least 1.97 $M_\odot$. All models are in agreement with AC2. On the other hand, the speculative PSR J1311-3430~\cite{Romani} is described only by few parametrizations: $\alpha_v = 0.0$ with GM1 and $\alpha_v \le 0.75$ with GM1LM are able to explain such high mass. Nevertheless,  all parametrization sets that can explain a pulsar with 2.1 $M_\odot$
 yield a very stiff EoS, in disagreement with  expected low density constraints
~\cite{Daniel}.

In ref.~\cite{Weiss2}, the authors propose a linear relation between the maximum mass and the strangeness fraction. 
In this work, as we see from Table~\ref{T7}, this relation is not present. However, a very interesting result turns up:
 all the parametrizations that are in agreement with the experimental constraint EC3, obtained from ref.~\cite{Daniel},
 yield exactly the same strangeness fraction of 0.206 in GM1 model. In GM1LM just one agrees with the EC3, the SU(6) parametrization set,
 which produces a value of $f_s= 0.168$.

It is also interesting to note that the trick of adding a new repulsive meson $\phi$ to stiffen the EoS indeed reproduces more massive neutron stars.
 However, the price we pay is that most of the parametrizations with GM1LM present very high pressure, which is in conflict with EC3. 
 As we have already said, the SU(6) is the only parametrization that conciliates EC3 with GM1LM. Moreover, observing  Fig.~\ref{F8} (right),
 we see that the SU(6) parametrization is in the upper limit of the pressure that remains in accordance with EC3. This strongly constraints
 the maximum allowed mass value around 2.06 $M_\odot$, at least in mean field QHD based models with our choice of couplings.  
As pointed out  earlier, the best parametrization that describes the experimental constraints is the GM3. Without hyperons, this parameterization
 predicts a neutron star with a mass of 2.04 $M_\odot$, in agreement  with all constraints analyzed, at all densities.
  Nevertheless, this parametrization rules out hyperon in neutron star cores, since no value of $\alpha_v$ is able to predict
 hyperonic neutron stars with masses larger than 1.91 $M_\odot$ as show in Fig.~\ref{F10}.

\begin{figure*}[ht]
\begin{tabular}{cc}
\includegraphics[width=5.6cm,height=6.2cm,angle=270]{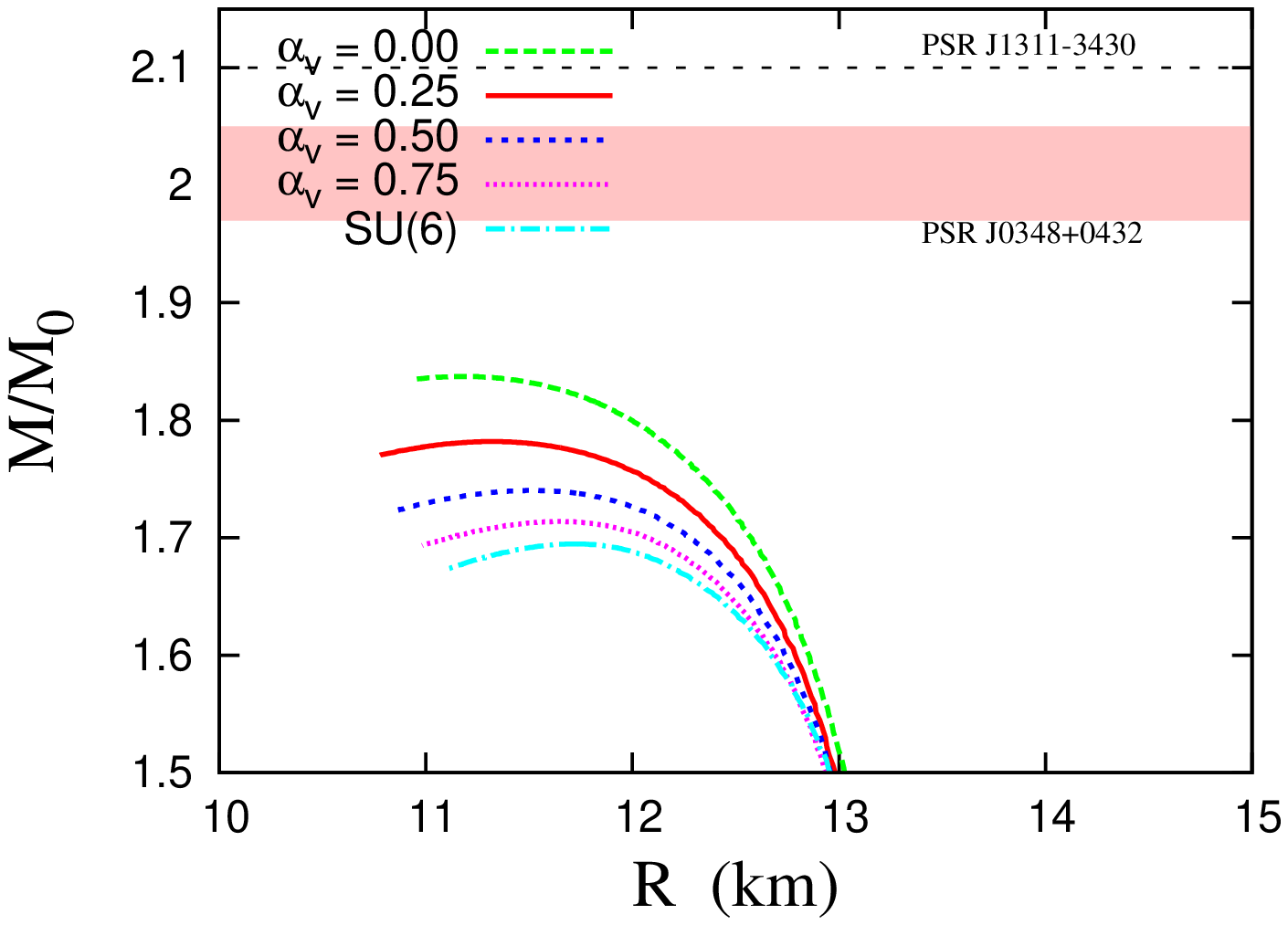} &
\includegraphics[width=5.6cm,height=6.2cm,angle=270]{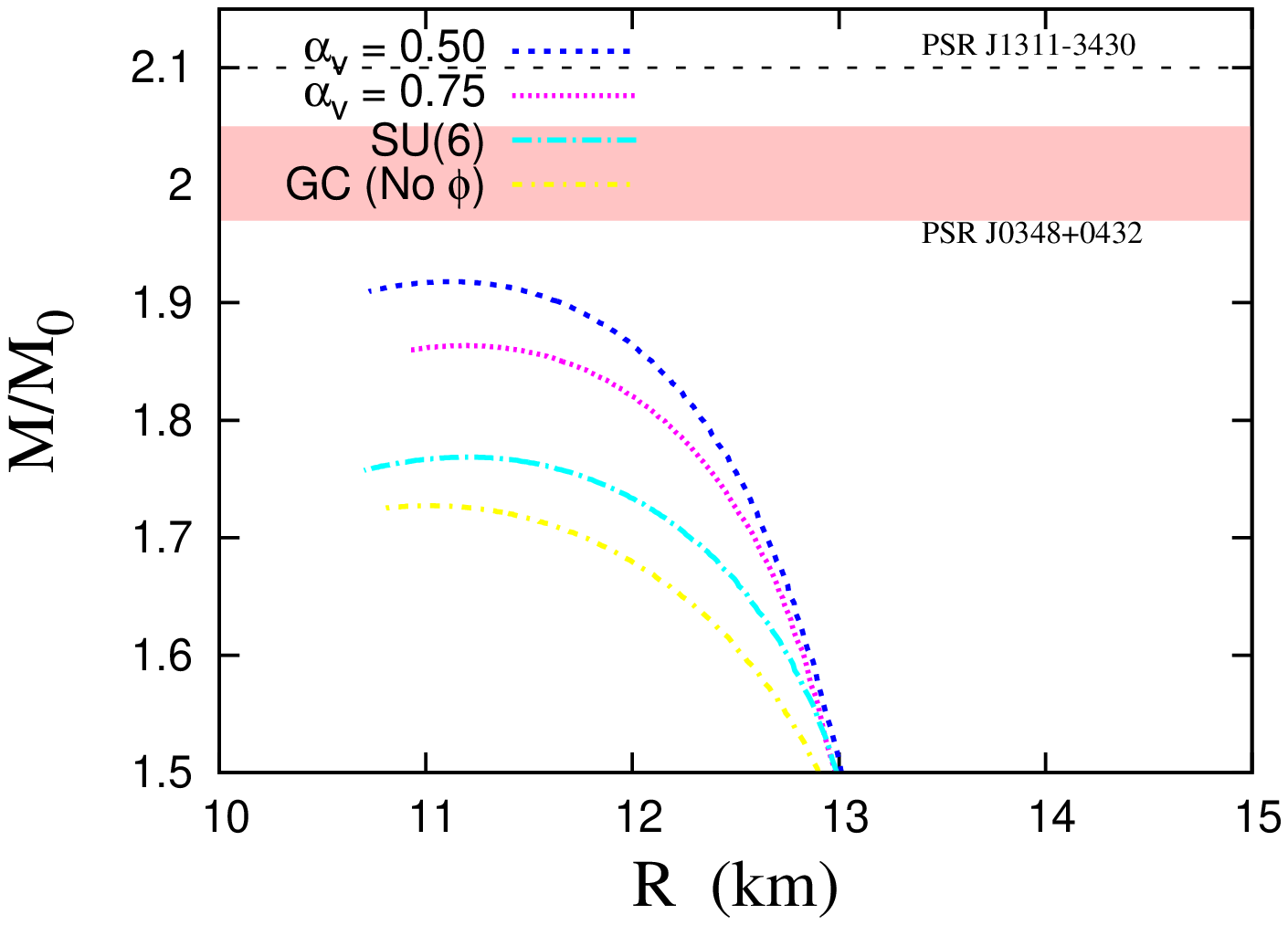} \\
\end{tabular}
\caption{Neutron stars mass/radii relation for GM3 (left) and GM3LM (right). No parametrization is able to predict the mass of PSR J0438+0432.} \label{F10}
\end{figure*}

The third astrophysical constraint (AC3) is a theoretical one. Based on chiral effective theory, ref.~\cite{Hebeler} constrains the radii of the
 canonical $1.4M_\odot$ neutron star to 9.7-13.9 $Km$. We plot in Table~\ref{T7b} the minimum and the maximum radii of the canonical mass for different
 models and parametrizations.

\begin{table}[ht]
\centering
\begin{tabular}{|c|c|c|}
\hline
 Model & Min./Max. radii ($1.4M_\odot$). &  AC3  \\
\hline
GM1,GM1LM & 13.73/13.85 Km & OK    \\
\hline
GM3,GM3LM & 13.06/13.14 Km & OK   \\
\hline
NL3 & 14.71 Km & Failed \\
\hline   
\end{tabular}
 \caption{Minimum and maximum radii of the canonical $1.4M_\odot$ for different models and parametrizations}\label{T7b}
 \end{table}
 
 We see that although close to the theoretical limit, the  GM1 and GM1LM agree with ref.~\cite{Hebeler}.
 GM3 and GM3LM also agree with AC3, but as said before, are in disagreement with  AC1 if hyperons are present. NL3 agrees with AC1 and AC2 but
 fails again when confronted with AC3.

It is worth bearing in mind that the measurement and assessment 
of neutron star radii still remain to be better understood. Recently, two different 
analyses of five quiescent low-mass X-ray binaries in 
globular clusters resulted in different ranges for neutron star radii.
While one of them, in which it was assumed that all neutron stars have the same 
radii, predicted that they should lie in the
range $R=9.1^{+1,3}_{1.5}$ \cite{guillot}, another calculation, based on a Bayesian 
analysis, foresees radii of all neutron stars to lie in between 10.9 and 12.7 Km~\cite{Lattimer2013}, 
which would put all previsions of our models in conflict with the experimental data
for the canonical 1.4$M_\odot$ neutron stars. However as pointed out by the authors,
better X-ray data is needed to
determine the compositions of accreting neutron
stars, as this can make 30\%
or greater changes in inferred neutron star radii~\cite{Lattimer2013}.


 Another astrophysical constraint is related to the maximum possible neutron star mass.
 According to \cite{Ruffini}, it must be lower than 3.2$M_\odot$ independently of the choice of the EoS.
  This theoretical constraint was updated in \cite{JSB}, where the maximum mass around 3.0$M_\odot$ was found due to the weakly repulsive nucleon potential. In our work the maximum mass is obtained with NL3 without hyperons and is equal to  $2.81M_\odot$. So all our models are in agreement with the maximum possible neutron star masses~\cite{JSB,Ruffini}.

Finally, we resume our main results in Table~\ref{T8}, indicating the way each parametrization behaves when confronted  with the six constraints used in this work. The  concept {\bf Fair} that appears
 in Table~\ref{T8} is due to the fact that the parametrization agrees with the EC3 at high densities, but continues to fail at low densities as pointed in Fig.~\ref{F7}. 

\begin{table}[ht]
\begin{tabular}{|c|c|c|c|c|c|c|c|}
\hline
 Model & $\alpha_v$ & EC1 &  EC2 & EC3 & AC1  & AC2 & AC3 \\
\hline
GM1 & No hyp. & OK  & OK & Failed & OK & OK & OK \\
\hline
 GM1 & SU(6)& OK   & OK  & Fair  &  Failed & OK  & OK \\
\hline
 GM1 & 0.75 & OK  & OK & Fair & OK & OK & OK\\
 \hline
GM1 & 0.50 & OK   &  OK & Fair & OK & OK & OK\\
\hline
GM1 & 0.25 & OK  & OK  & Failed & OK & OK & OK  \\
\hline
GM1 &  0.0 & OK  & OK & Failed  & OK & OK & OK   \\
\hline
GM1 & GC & OK  & OK & Failed &  OK  & OK & OK \\
\hline
GM1LM & SU(6) & OK  & OK & Fair & OK & OK & OK  \\
\hline
GM1LM & 0.75 & OK    & OK  & Failed  & OK & OK & OK   \\
\hline
GM1LM & 0.50 & OK   & OK & Failed &  OK  & OK & OK \\
\hline
GM3 & No hyp. & OK  & OK & OK & OK & OK & OK \\
\hline
GM3 & All/GC  & OK  & OK & OK & Failed& OK  & OK \\
\hline
GM3LM & All  & OK  & OK & OK & Failed& OK & OK  \\
\hline
NL3 & No hyp.  & Failed  & Failed & Failed & OK & OK & Failed  \\
\hline
\end{tabular}
 \caption{Our choice of the parametrizations when confronted with experimental and observational constraints.}\label{T8}
 \end{table}

\section{Conclusion \label{sec5}}

In this work we investigated the hyperon onset in hypernuclear matter imposing a complete  SU(3) symmetric model, where we propose that all hyperon-meson
 couplings need to obey the SU(3) symmetry properties in order to reduce the number of free parameters of the theory. Utilizing a QHD based model,
 we analyze three widely used parametrizations:  GM1, GM3 and NL3 in two different models, that we refer to as $\sigma\omega\rho$  and $\sigma\omega\rho\phi$.
 We then test them against experimental and astrophysical constraints obtained in the last decade. We see that NL3, although describes
  nuclear matter properties very well, fails to describe dense asymmetric matter. Also, although we can predict very massive hyperonic stars with many of
 the investigated parametrizations, a maximum mass of 2.06 $M_\odot$ arises in order to describe the soft EoS at low densities regime.
 This constraint prevents us from explaining the mass of the speculative PSR J1311-3430 since none of the parametrizations that agree with EC3 can
 reproduce such high mass. 

 With the GM1LM, the trick of adding a new vector meson in order to stiff the EoS seems to be valid only if the SU(6) choice of couplings is used.
 Lower values of $\alpha_v$ enters in conflict with EC3. Also, if we assume that GM1 is a good parametrization to describe  nuclear properties,
 this implies that the hyperon production is not only possible, but necessary to soften the EoS and reconcile theory and experience.

We also analyze some theoretical features of the model, as the speed of sound in 
dense nuclear matter. We see that stiffer EoS have also higher value of $v_s$. 
Since there exists a theoretical limit for the speed of sound in quark matter,
  a measurement of this physical quantity may be important to rule out quark-hadron phase transitions at specific densities. Also, an unexpected relation arises
when we compare the speed of sound with the EC3, what ultimately  constraints  the hyperon onset.

The possibility of a linear relation between the maximum stellar mass and the 
strangeness fraction found in \cite{Weiss2} was also investigated for the present 
choice of parameters.
Within our prescription,  we do not see this relation. 

The role of the hyperon potential as proposed in~\cite{Weiss} was also checked and 
we found that less repulsive potentials produce stiffer EoS. This is due to the 
fact that the vector meson channel dominates at high densities, and the role of 
hyperon potentials plays only a secondary role.

Despite all the efforts made in recent years toward a better understand of nuclear matter
and its implication in neutron stars properties, there is still much work to be done.
The possibility of a hypermassive 2.7$M_\odot$ neutron star~\cite{Romani}, and very compact ones~\cite{guillot,Lattimer2013}
are examples of still unexplained phenomena.

The consequences of considering  the scalar-isovectorial $\delta$ meson are the 
next step of the present work. 
Also, as pointed in the literature~\cite{Lopes2,Dexheimer}, effects of strong 
magnetic field could be important in the description of magnetars. Works along 
these lines are in progress.

\begin{acknowledgments}

 This work was partially supported by CAPES, CNPq and FAPESC. We would like 
to thank Dr. Luis B. Castro for his patience in checking some Fortran codes 
and for pointing out some ambiguities in the manuscript. 

\end{acknowledgments}

\appendix*

\section{SU(3) symmetry group}

 In order to consider a completely symmetric theory for the strong interaction based on a QHD model, we impose that the Yukawa type interaction is invariant under SU(3) transformations. In what follows we just consider the electric coupling since the magnetic coupling does not contribute in a mean field approximation \cite{Swart2}. The Yukawa interaction is expressed as \cite{Swart}:
 
 \begin{equation}
 \mathcal{L}_{YUK} = -g(\bar{\psi}_B\psi_B)M , \label{a1}
 \end{equation}
 where $\psi_B$ is the Dirac field of baryons and $M$ is the field of an arbitrary meson. This Lagrangian belongs to the irreducible representation IR\{1\},
a unitary singlet. All the baryons of the model we consider belong to the IR\{8\}. The mesons of the vector IR\{8\} are the $\omega_8$ and $\rho^0$.  $\phi_1$ is a vector singlet, $\sigma_8$ belongs to the IR\{8\} of the scalar octet and $\sigma_1$ is a scalar singlet. Now, to preserve the unitary symmetry, $(\bar{\psi}_B\psi_B)$ must transform as:
 
 \begin{itemize}
 
 \item IR\{8\} when $M$ belongs to  IR\{8\},
 \item IR\{1\} when $M$ belongs to  IR\{1\}.
 
 \end{itemize}
 
 However, by the Speiser method~\cite{Swart}, there are two ways to couple $\{8\} \otimes \{8\}$ to $\{8\}$, typically the symmetric and the antisymmetric ones~\cite{Stancu}.  Therefore, the Yukawa interaction can be written as:
 
 \begin{eqnarray}
 \mathcal{L}_{YUK} = -(g_1^8 \mathcal{C}^{1} + g_2^8 \mathcal{C}^{2})(\bar{\psi}_B\psi_B)M, \\ 
\quad \mbox{} \nonumber , \label{a2}
 \end{eqnarray}
for the mesons belonging to $IR\{8\}$, and 

 \begin{eqnarray}
 \mathcal{L}_{YUK} = -(g^1)M,  \label{a3}
 \end{eqnarray}
for the mesons belonging to $IR\{1\}$, 
 where  $\mathcal{C}^1$ and $\mathcal{C}^2$ are the SU(3) Clebsch-Gordan (CG) coefficients of the symmetric and antisymmetric coupling respectively. In this work we use the CG as in ref.~\cite{Mc}. Following ref.~\cite{Swart} we introduce the constants:
 
 \begin{eqnarray}
 g_8 = [\sqrt{30}/40g_1^8 + (\sqrt{6}/24)g_2^8], \quad \mbox{and,} \nonumber \\ \quad \alpha = (\sqrt{6}/24)(g_2^8/g_8), \quad \; \;  \label{a4}
 \end{eqnarray}
 what allow us to write the coupling constants of the baryons with the vector mesons as:
\begin{eqnarray}
g_{NN\rho} = g_{8v}, \quad g_{\Sigma\Sigma\rho} = 2g_{8v}\alpha_v, \quad \nonumber \\
 g_{\Xi\Xi\rho} = -g_{8v}(1 -  2\alpha_v), \quad  g_{\Lambda\Lambda\rho} = 0, \nonumber \\
g_{NN\omega_8} =\frac{1}{3}g_{8v}\sqrt{3}(4\alpha_v -1), \nonumber \\
 g_{\Sigma\Sigma\omega_8} = \frac{2}{3}g_{8v}\sqrt{3}(1 - \alpha_v), \nonumber \\   g_{\Xi\Xi\omega_8} = -\frac{1}{3}\sqrt{3}g_{8v}(1 + 2\alpha_v), \nonumber \\
g_{\Lambda\Lambda\omega_8}  = - \frac{2}{3}g_{8v}\sqrt{3}(1 - \alpha_v), \nonumber \\ 
 g_{NN\phi_1} =  g_{\Sigma\Sigma\phi_1} = g_{\Lambda\Lambda\phi_1} = g_{\Xi\Xi\phi_1} = g_{1v}  , \label{a5}
\end{eqnarray}
while the couplings with the scalar ones read:
\begin{eqnarray}
g_{NN\sigma_8} =\frac{1}{3}g_{8s}\sqrt{3}(4\alpha_s -1), \nonumber \\
 g_{\Sigma\Sigma\sigma_8} = \frac{2}{3}g_{8s}\sqrt{3}(1 - \alpha_s), \nonumber \\
  \quad  g_{\Xi\Xi\sigma_8} = -\frac{1}{3}\sqrt{3}g_{8s}(1 + 2\alpha_s), \nonumber \\
g_{\Lambda\Lambda\sigma_8}  = - \frac{2}{3}g_{8s}\sqrt{3}(1 - \alpha_s), \nonumber \\ 
g_{NN\sigma_1} =  g_{\Sigma\Sigma\sigma_1} = g_{\Lambda\Lambda\sigma_1} = g_{\Xi\Xi\sigma_1} = g_{1s}  , \label{a6}
\end{eqnarray}
where the new subscripts $v$ and $s$ appear to differentiate the set of vector mesons from the set of the scalar ones.
 Also, all the constructions that couple baryons of different species are ignored.

In nature, the observed $\omega$ and $\phi$ mesons are not the theoretical $\omega_8$ and $\phi_1$ ones, but a mixture of them~\cite{Dover}.
So, the coupling constants of the real vector mesons with the baryons read:
\begin{eqnarray}
g_{NN\omega} =  \cos\theta_v g_{1v} + \sin\theta_v \frac{1}{3}\sqrt{3}g_{8v}(4\alpha_v - 1), \nonumber \\
g_{\Sigma\Sigma\omega} =  \cos\theta_v g_{v1} + \sin\theta_v \frac{2}{3}\sqrt{3}g_{8v}(1 - \alpha_v ), \nonumber \\
g_{\Lambda\Lambda\omega} =  \cos\theta_v g_{1v} - \sin\theta_v \frac{2}{3}\sqrt{3}g_{8v}(1 - \alpha_v ), \nonumber \\
g_{\Xi\Xi\omega} =  \cos\theta_v g_{1v} - \sin\theta_v \frac{2}{3}\sqrt{3}g_{8v}(1 + 2\alpha_v ). \label{a7}
\end{eqnarray}

The results for the $\phi$ coupling are similar, just replacing $\cos \theta_v \to -\sin\theta_v$ and $\sin \theta_v \to \cos \theta_v$~\cite{Dover}.
In the case of the real scalar mesons $\epsilon(760)$ (what we call here just $\sigma$) and the $f_0(980)$ (call here $\sigma^*$)~\cite{Greiner} the procedure is entirely analogous to the $\omega$ and $\phi$ respectively, just replacing $\omega$ by $\sigma$, $\phi$ by $\sigma^*$ and  the subscript $v$ by $s$ in Eq. (\ref{a7}). In order to obtain stiffer EoS the $\sigma^*$ meson is not  considered as in \cite{Weiss,Weiss2}.

Since the meson nucleon parametrization is fixed, we have a priori six free parameters: $z_v =(g_{8v}/g_{1v})$, $\theta_v$, $\alpha_v$, $z_s = (g_{8s}/g_{1s})$, $\theta_s$ and $\alpha_s$. To reduce the numbers of free parameters, for the vector mesons, we use the hybrid SU(6) symmetry group~\cite{Greiner,Stancu} to fix the $z_v = \sqrt{6}$ and $\theta_v = 35.264$, what correspond to an ideal mixing angle. This leaves just $\alpha_v$ as free parameter. When $\alpha_v = 1$, we recover the complete SU(6) parametrization for the vector mesons coupling~\cite{Pais}. This approach based on the SU(6) symmetry for the vector mesons is widely accepted in the literature~\cite{Weiss,Weiss2,Swart2,Dover,Swart3,Swart4}. 

On the other hand, for the scalar mesons the literature is  controversial. In ref.~\cite{Greiner}  the $\sigma$ meson is considered a  member of the IR\{8\}, while in \cite{Swart3,Swart4} it is considered as a true member of the IR\{1\}. In \cite{Carr} , the $\sigma$ meson is taken as the mixing of not only two, but three scalar mesons. Finally, in \cite{Rijken,Swart5} an almost ideal mixing is assumed. We follow the last prescription and consider $\theta_s$ = 35.254 as an ideal mixing approximation. For the $z_s$, we use a {\it near SU(6) symmetry}  with $z_s =\frac{8}{9}\sqrt{6}$. The reason for choosing 
{\it near SU(6) symmetry} parametrization is because we have checked that SU(6) fixes $z_s = \sqrt{6}$, what results in a very repulsive $\Xi$ potential $U_\Xi = +60 MeV$, in disagreement with expected values running from $-40$ to $40$ MeV \cite{Weiss}.  Then, only $\alpha_s$  remains to be fixed.
Now we attach $\alpha_s$ to  $\alpha_v$ forcing the $U_\Lambda$ potential to  be equal to -28 MeV~\cite{Glen2}.
 This leaves our theory in complete tune with the SU(3) symmetry group for all hyperon-mesons couplings,
 and only one free parameter is left to be varied, $\alpha_v$.

\end{document}